\pdfoutput=1
\documentclass[journal]{IEEEtran}

\usepackage{cite}

\ifCLASSINFOpdf
  \usepackage[pdftex]{graphicx}

  \DeclareGraphicsExtensions{.pdf,.jpeg,.png}
\else

\fi
\ifCLASSOPTIONcompsoc
\usepackage[caption=false,font=normalsize,labelfon
t=sf,textfont=sf]{subfig}
\else
\usepackage[caption=false,font=footnotesize]{subfi
g}
\fi
\usepackage{amsmath}
\usepackage{stix}
\usepackage{threeparttable}
\usepackage{array}
\usepackage{xcolor}
\usepackage{soul}
\soulregister\cite7

\usepackage{pdfpages}
\usepackage{tikz}
\usetikzlibrary{calc,shapes,arrows,chains,arrows.meta}
\usepackage[export]{adjustbox}
\usepackage{upgreek}
\definecolor{newTextColor}{RGB}{0,0,0} 
\definecolor{modTextColor}{RGB}{0,0,0} 
\definecolor{typoColor}{RGB}{0,0,0}

\newcounter{node}
\usepackage[hidelinks]{hyperref}
\makeatletter
\newcommand{\customlabel}[2]{%
  \protected@write \@auxout {}{\string \newlabel {#1}{{#2}{\thepage}{#2}{#1}{}}}%
  \hypertarget{#1}{#2}
}
\tikzset{autonumbered node/.style={/utils/exec={\stepcounter{node}}},label=above:{\alph{node}}}
\newcommand{\nodelabel}[1]{\protect\customlabel{#1}{\alph{node}}}
\makeatother

\begin{document}

\onecolumn
\pagenumbering{Roman}
\textbf{Author's pre-print}
\newline\newline
This article has been accepted for publication in IEEE Transactions on Antennas and Propagation. This is the author's version which has not been fully edited and
content may change prior to final publication. Citation information: DOI~10.1109/TAP.2023.3291424

$\copyright$ 2023 IEEE. Personal use of this material is permitted. Permission
from IEEE must be obtained for all other uses, in any current or future
media, including reprinting/republishing this material for advertising or
promotional purposes, creating new collective works, for resale or
redistribution to servers or lists, or reuse of any copyrighted
component of this work in other works. See https://www.ieee.org/publications/rights/index.html for more information.
\newpage
\twocolumn
\pagenumbering{arabic} 
%
\title{Evaluation Method and Design Guidance \\ for Direction Finding Antenna Systems}
%
%
%
\author{Lukas~Grundmann, \IEEEmembership{Graduate Student Member, IEEE} and  Dirk~Manteuffel, \IEEEmembership{Member, IEEE}
\thanks{This work is performed in the project Master360 under grant 20D1905C, funded by the German Federal Ministry for Economic Affairs and Climate Action within the Luftfahrtforschungsprogramm (LuFo).
}
\thanks{The authors are with the Institute of Microwave and Wireless Systems, Leibniz University Hannover, Appelstr. 9A, 30167 Hannover, Germany \mbox{(e-mail:} \mbox{grundmann@imw.uni-hannover.de;} \mbox{manteuffel@imw.uni-hannover.de)}
}}

\maketitle

\begin{abstract}
A deterministic evaluation procedure for multi-port direction finding antennas is proposed. 
It is based on a direction finding uncertainty parameter, which describes how well different directions of arrival and polarizations are distinguishable. 
By investigating a simple antenna array, it is shown that the proposed parameter provides additional insight into the behavior of an antenna system, when compared to established methods. 
Moreover, since the uncertainty parameter is calculated from a set of far fields, it is applicable to port far fields as well as Characteristic Modes. 
This finding is utilized to derive a design guidance: 
Starting with a set of Characteristic Mode far fields, the angular distribution of the uncertainty is investigated to verify that no ambiguities are present. 
Different sets of far fields are compared and the differences regarding their direction finding behavior are visualized and explained using the uncertainty in conjunction with an estimate of the incident field. 
To quantify these differences, a key performance indicator is introduced that summarizes the direction finding capabilities over a selected angular region. 
To demonstrate the design process, a {\color{typoColor}multi-mode multi-port antenna} with three uncorrelated ports is developed, manufactured and measured. 
\end{abstract}

\begin{IEEEkeywords}
direction finding antennas, characteristic modes, multi port antennas, direction of arrival
\end{IEEEkeywords}

%
\IEEEpeerreviewmaketitle

\section{Introduction} \label{sec:Introduction}

\begin{figure}
    \centering
    \includegraphics{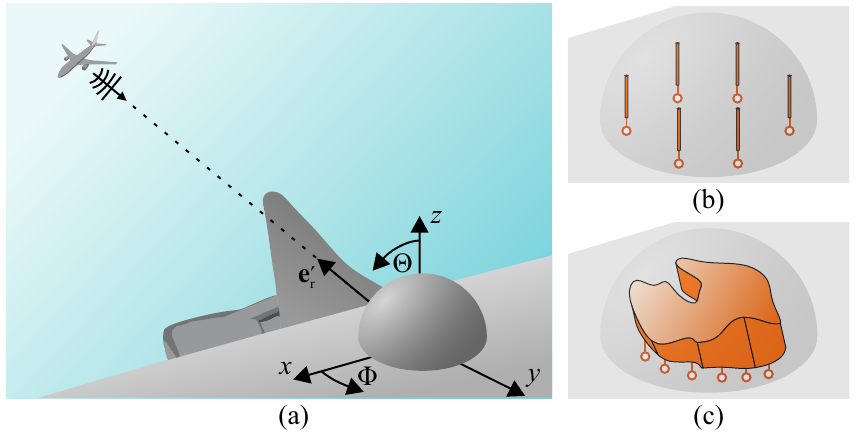}
    \caption{Aerial direction finding scenario (a). The multi-port antenna system is positioned under the hemispherical radome on an aircraft hull. For example, it can comprise several antennas in an array (b), or a connected multi-port antenna structure (c). The DoA of a plane wave originating from a second aircraft shall be estimated.}
    \label{fig:introduction}
\end{figure}

\IEEEPARstart{D}{irection} finding (DF) is one of the major applications for multi-port antenna systems. It is required for e.\,g. communication applications like user tracking by 5G base stations \cite{Wild2021}. Also, safety features like the airborne collision avoidance system (ACAS) rely on direction finding to determine the relative position of other traffic participants \cite{ICAO2006}, see Fig.~\ref{fig:introduction}a. 
{\color{typoColor}Most} implementations of multi-port direction finding systems are based on antenna arrays, like the one depicted in Fig.~\ref{fig:introduction}b. Here, the analytic relations between the signals at the antenna ports are utilized to determine the direction of arrival (DoA) of an incident plane wave \cite{Balanis2016}. Alternatively, the multi-port antenna system can employ a single conducting structure, see Fig.~\ref{fig:introduction}c. Using Characteristic Modes (CM) \cite{Harrington1971a} to design these multi-mode multi-port antennas (M$^3$PA) is an established approach for communication applications \cite{Peitzmeier2019, Peitzmeier2022, Manteuffel2022}. Moreover, M$^3$PA recently received increased attention for direction finding applications \cite{Ma2019, Poehlmann2019, AlkubtiAlmasri2019, Ren2021, Grundmann2021, Grundmann2022}. 

Since the analytic relationships between the DoA and the port signals used for antenna arrays are no longer valid for M$^3$PA, established methods for the direction finding have to be adapted, as was done in \cite{Poehlmann2019}. Crucially, a method to compare the expected performance of different antenna systems is required. 
The most common evaluation procedures are based on the Cramer Rao Bound (CRB), e.\,g. \cite{Weiss1991, Hua1991, Nordebo2006, Jackson2015, Pralon2017, Poehlmann2019, Ma2019, Li2020, Ma2022}. 
The CRB is a statistical parameter used to predict the performance of a direction finding system, assuming an ideal estimation algorithm. The most widespread DoA estimation algorithms are the multiple signal classification (MUSIC) \cite{Schmidt1986} and related methods. In e.\,g. \cite{DeB.Gripp2017, Kabiri2020, Park2021, Ma2022} the algorithms' results, the so called MUSIC spectra, are used to illustrate the performance of the antenna system for specific DoAs. In e.\,g. \cite{Jackson2015, Kataria2019}, the estimation error of the MUSIC algorithm is given to evaluate an antenna system. All of these methods are suited tools to quantify the performance of a direction finding system, including the influence of the algorithm, noise, the number of samples w.\,r.\,t. time and the number of simultaneous signals. For antenna development however, using these statistical and algorithmic parameters introduces unnecessary levels of complexity. These obstruct the understanding of the behavior of the antenna that is required to refine the antenna system. In particular, if noise is neglected, no useful information for the antenna development is provided by the established methods. 

Therefore, an evaluation method is required that is based only on deterministic antenna properties, providing additional insight for the antenna design process. {\color{newTextColor} First steps towards this goal are found in \cite{Grundmann2021}, where we present a concept to determine the DoA directly from CM currents. In \cite{Grundmann2021a}, estimated incident fields for direction finding are added as a tool for visualization and interpretation. Furthermore, the concept of a selection method for CM based on these findings is outlined in \cite{Grundmann2022}. However, these works lack a rigorous derivation that considers the signal power and allows the application to both ports and CM. This is required for a comparison of the deterministic evaluation method to established statistical methods and the design and manufacturing of a demonstrator.}
{\color{modTextColor} Therefore, in this work, the vector correlation direction finding method \cite{Bailey2012} is used as a starting point for the derivation. It is outperformed by MUSIC in terms of DoA estimation accuracy, but the results only depend on the antenna patterns. As these are available for both ports and CM, CM far fields are used in early stages of the antenna design process. Findings from these investigations are transferred to port far fields later on.}
To this end, Sec.~\ref{sec:method} introduces the uncertainty parameter, estimated incident fields and a key performance indicator (KPI) to evaluate and interpret the direction finding behavior of an antenna system. Sec.~\ref{sec:examplesEvaluation} provides examples for the evaluation of antenna setups and a comparison to the established methods. Finally, Sec.~\ref{sec:exampleDesign} gives an example for an antenna design process using the proposed method, including the fabrication and measurement of a demonstrator. 

\section{Evaluation of Direction Finding Antennas} \label{sec:method}
\subsection{General Description} \label{sec:generalDescription}
First, a general description of a $P$-port direction finding antenna system is required. 
A total of $K$ samples of the radiated far field $F_{\gamma,p}(\mathbf{e}_{\mathrm{r},k})$ of each antenna port $p$ are determined by measurement or simulation. Each sample $k$ belongs to a specific direction $\mathbf{e}_{\mathrm{r},k}$ and polarization $\gamma_k$. For simplicity, the index $k$ of the polarization is omitted in the following. The complex far field values are related to the far field approximation of the electric field in a distance $r$ from the coordinate origin by \cite{Capek2023}
\begin{equation} \label{eq:FFdefinition}
    E_{\gamma,p}(\mathbf{e}_{\mathrm{r},k}) = 
    F_{\gamma,p}(\mathbf{e}_{\mathrm{r},k}) \frac{\mathrm{e}^{-\mathrm{j}k_0r}}{r} \,,
\end{equation}
where $\mathrm{j}$ is the imaginary unit and $k_0$ the free space wave number. 

\begin{figure}
    \centering
    \includegraphics{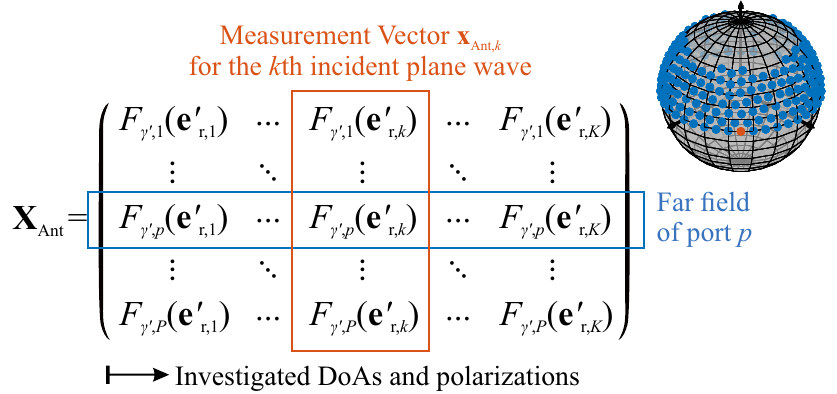}
    \caption{Measurement matrix $\mathbf{X}_\mathrm{Ant}$. An example for a set of investigated DoAs is shown in the top right hand corner. }
    \label{fig:FarFieldMeasMat}
\end{figure}
As illustrated in Fig.~\ref{fig:introduction}a, a plane wave with some amplitude and phase is incident onto the antenna system from DoA $\mathbf{e}'_{\mathrm{r},k}$ with polarization $\gamma'$. Throughout this work, primed quantities are used to indicate an incident wave. 
The signal measured at each port is proportional to the far field $F_{\gamma',p}(\mathbf{e}'_{\mathrm{r},k})${\color{newTextColor}, see (\ref{eq:VfromK}) in the appendix}. These far field values are assembled in a $P$-by-$1$ measurement vector $\mathbf{x}_{\mathrm{Ant},k}$\footnote{Since noise is neglected and unit signal power is assumed, this is equivalent to the steering vector or antenna response vector used in the literature, e.\,g. \cite{Poehlmann2019}.}. This is repeated for all $K$ sample DoAs. The resulting measurement vectors are assembled in the $P$-by-$K$ matrix $\mathbf{X}_\mathrm{Ant}$, depicted in Fig.~\ref{fig:FarFieldMeasMat}. 
If only a single polarization $\gamma'$ is considered, and the DoAs cover all points on a spherical grid, each row of $\mathbf{X}_\mathrm{Ant}$ contains the sampled, $\gamma'$-polarized far field pattern of the respective port. 

In order to identify requirements for a direction finding antenna system, the properties of $\mathbf{X}_\mathrm{Ant}$ are investigated in the following. 
Firstly, each DoA should result in a unique measurement vector $\mathbf{x}_{\mathrm{Ant,}k}$ that is decorrelated from other columns of $\mathbf{X}_\mathrm{Ant}$. The correlation between the $\alpha$th and $\beta$th column of $\mathbf{X}_\mathrm{Ant}$ is \cite{Bailey2012}: 
\begin{equation} \label{eq:DoACorrelations}
    \rho_{\alpha,\beta} = \frac{\mathbf{x}_\mathrm{Ant,\alpha} ^\mathrm{H} \mathbf{x}_\mathrm{Ant,\beta}}{||\mathbf{x}_\mathrm{Ant,\alpha}|| ||\mathbf{x}_\mathrm{Ant,\beta}||} \,,
\end{equation}
with $(\cdot)^\mathrm{H}$ being the conjugate transpose and $||\cdot||$ the euclidean vector norm. The number of possible DoAs $K$ is larger than the number of ports $P$ in almost any application, and even infinite in many. Consequently, the measurement vectors are correlated and the desired $\rho_{\alpha,\beta}=0$ with $\alpha\neq\beta$ is not fulfilled for all DoAs. Still, a low correlation is desired for a well suited DF antenna system, in particular for DoAs that are further apart from each other. 

For an ideal direction finding system, {\color{typoColor}a constantly high received power is desired} for all considered DoAs. In order to evaluate DF antennas, the influence of the received power on the direction finding performance is combined with the correlation to give
\begin{equation} \label{eq:rhoNormalized}
    u_{\alpha,\beta} = \frac{\rho_{\alpha,\beta} }{||\mathbf{x}_\mathrm{Ant,\alpha}|| ||\mathbf{x}_\mathrm{Ant,\beta}||} \,,
\end{equation}
which we call the uncertainty parameter. This way, higher directivities in the investigated directions are rewarded with a lower uncertainty parameter. Since in contrast to $\rho_{\alpha,\beta}$, $u_{\alpha,\beta}$ depends on the magnitude of the far field, it varies w.\,r.\,t. the normalization. Depending on whether the far field defined by (\ref{eq:FFdefinition}) is derived from the directivity, the gain or the realized gain, $u_{\alpha,\beta}$ includes the influence of the respective underlying physical effects. 

These findings also have implications for the rows of $\mathbf{X}_\mathrm{Ant}$ and thereby the complex far fields, see Fig.~\ref{fig:FarFieldMeasMat}. For any two ports, these must differ from each other, in polarization, magnitude or phase w.\,r.\,t. the coordinate origin. This is evident since nearly identical ports do not provide additional information, in particular if noise is considered \cite{Kataria2019}. 
Additionally, the coupling between antenna ports, and thereby the correlation of the port far fields, is one of the main design concerns. If the coupling of the ports is high, the active impedance \cite{Pozar2003} of each port depends on the DoA. Due to the resulting mismatch, the total received power is reduced, increasing the uncertainty (\ref{eq:rhoNormalized}). 
Although the far field correlations cannot replace (\ref{eq:DoACorrelations}), because the need to cover all selected DoAs is not inherently considered, uncorrelated ports are still beneficial. 
For the design of such ports, Characteristic Modes are a helpful tool since they describe an antenna structure in terms of orthogonal far fields.

\subsection{Application to Characteristic Modes} \label{sec:CharacteristicModes}
\tikzstyle{line} = [draw, -latex']
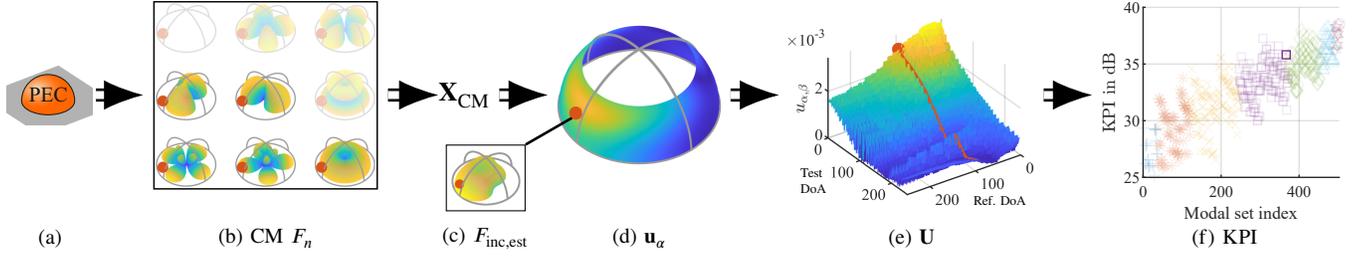
\begin{figure*} 
\begin{tikzpicture}

\node[above left] (imgOpt) at (\textwidth-4.5pt,0) {  
    \includegraphics[width=0.18\textwidth] {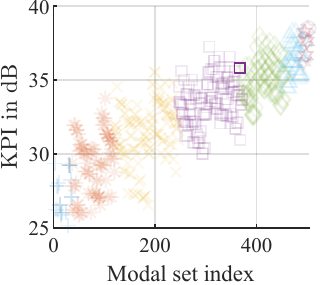}
};

\node[above = 0.1 of imgOpt.center] (arrowHeight) {};

\node[left = 0.65 of imgOpt] (imgCprime) {  
    \includegraphics[width=0.18\textwidth] {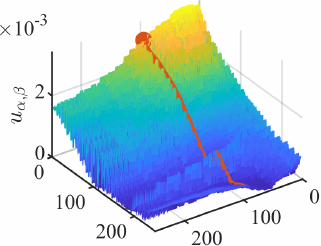}
};
\node[below left = 0.6 and 0.8 of imgCprime.center] (tstDoACprime) {\tiny \begin{tabular}{l} Test\\ DoA\end{tabular}};
\node[below right = 1 and 0.7 of imgCprime.center] (refDoACprime) {\tiny Ref. DoA};

\node[left = 0.65 of imgCprime.west|-arrowHeight] (imgCcol) {
    \includegraphics[width=0.12\textwidth] {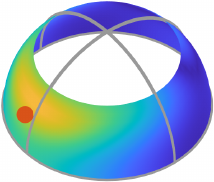}
};

\node[left = 0.65 of imgCcol] (XNode) {$\mathbf{X}_\mathrm{CM}$};

\node[left = 0.55 of XNode] (imgModes) {
    \includegraphics[width=3cm] {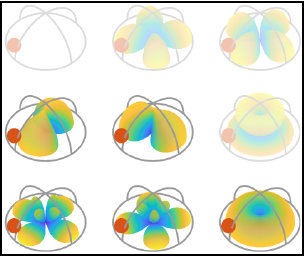}
};

\node[right, inner sep=0] (imgCAD) at (0,42|-arrowHeight) {
    \includegraphics[]{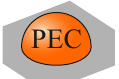}
};

\node[rectangle,draw,left = 0.3 of {imgCcol.west|-42,0.7}, inner sep=2] (imgEincRef) {
    \includegraphics[width=0.05\textwidth] {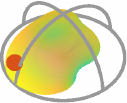}
};

\node[below left = 0.11 and 0.65 of imgCcol.center] (redDot){};
\draw [line width=1pt] (imgEincRef.east|-imgEincRef.north) -- (redDot);

\draw [line width=1pt, double distance=3pt,
        arrows = {-Latex[length=0pt 3 0]}] (imgCAD.east|-arrowHeight) -- (imgModes.west|-arrowHeight);
\node[right = 0 of XNode.west] (assistNode) {};
\draw [line width=1pt, double distance=3pt,
        arrows = {-Latex[length=0pt 3 0]}] (imgModes.east|-arrowHeight) -- (assistNode|-arrowHeight);
\draw [line width=1pt, double distance=3pt,
        arrows = {-Latex[length=0pt 3 0]}] (XNode.east|-arrowHeight) -- (imgCcol.west|-arrowHeight);
\draw [line width=1pt, double distance=3pt,
        arrows = {-Latex[length=0pt 3 0]}] (imgCcol.east|-arrowHeight) -- (imgCprime.west|-arrowHeight);
\draw [line width=1pt, double distance=3pt,
        arrows = {-Latex[length=0pt 3 0]}] (imgCprime.east|-arrowHeight) -- (imgOpt.west|-arrowHeight);

\node[](labelHeight) at (0,-0.1) {};
\node[autonumbered node] (txtalpha)  at (imgCAD.south|-labelHeight) {\footnotesize $\mathrm{(\nodelabel{fig:OverviewStructure})}$};
\node[autonumbered node] (txtbeta) at (imgModes.south|-labelHeight) {\footnotesize $\mathrm{(\nodelabel{fig:OverviewFarFields})}$ CM $F_n$};
\node[autonumbered node] (txta) at (imgEincRef.south|-labelHeight) {\footnotesize $\mathrm{(\nodelabel{fig:OverviewEstInc})}~ F_\mathrm{inc,est}$};
\node[autonumbered node] (txtb) at (imgCcol.south|-labelHeight) {\footnotesize $\mathrm{(\nodelabel{fig:OverviewUvec})}~ \mathbf{u}_\alpha$};
\node[autonumbered node] (txtd) at (imgCprime.south|-labelHeight) {\footnotesize $\mathrm{(\nodelabel{fig:OverviewUmat})}~ \mathbf{U}$};
\node[autonumbered node] (txte) at (imgOpt.south|-labelHeight) {\footnotesize $\mathrm{(\nodelabel{fig:OverviewKPI})}~ \mathrm{KPI}$};
\end{tikzpicture}
\centering
\caption{Overview of the proposed evaluation and design guidance procedure. 
The CM far fields $F_n$ of an antenna structure under development (\ref{fig:OverviewStructure}) are shown in  (\ref{fig:OverviewFarFields}). From this, the non-transparent modes are used to form the measurement matrix $\mathbf{X}_\mathrm{CM}$. For a reference DoA $\alpha$, indicated by the red dot, the estimated incident field is visualized (\ref{fig:OverviewEstInc}) and the uncertainty to all other DoAs is determined and stored in a vector $\mathbf{u}_\alpha$ (\ref{fig:OverviewUvec}). The uncertainty matrix $\mathbf{U}$ provides a combined display for all reference DoAs (\ref{fig:OverviewUmat}). From this, the KPI for the selected set of modes is determined (\ref{fig:OverviewKPI}). It is compared to that of other sets of modes, represented by the pale colored data points in (\ref{fig:OverviewKPI}). 
\label{fig:Overview} }
\end{figure*}
Before Characteristic Modes are employed for direction finding, a brief recap of relevant properties of the CM is provided.
Characteristic Modes are determined using e.\,g. \cite{Harrington1971a} to yield a set of CM eigencurrents $\mathbf{I}_n$ with respective eigenvalues $\lambda_n$ and far fields $F_{\gamma,n} (\mathbf{e}_\mathrm{r})$. The far fields are superimposed to yield a total scattered far field 
\begin{equation} \label{eq:sctFnSuperposition}
    F_{\gamma',\gamma} (\mathbf{e}'_\mathrm{r},\mathbf{e}_\mathrm{r}) = \sum\limits_{n=1}^N \frac{1}{2} \frac{\mathbf{I}_n^\mathrm{H} \mathbf{V}_\mathrm{\gamma'} (\mathbf{e}_\mathrm{r}')} {1 + \mathrm{j} \lambda_n} F_{\gamma,n} (\mathbf{e}_\mathrm{r}) 
    = \sum\limits_{n=1}^N a_{\gamma',n}  (\mathbf{e}_{\mathrm{r},k}') F_{\gamma,n} (\mathbf{e}_\mathrm{r})  \,,
\end{equation}
where $\mathbf{V}_\mathrm{\gamma'} (\mathbf{e}_\mathrm{r}')$ is the incident electric field on the surface, defined in the appendix, and $a_{\gamma',n}  (\mathbf{e}_{\mathrm{r},k}')$ is the modal weighting coefficient. 
In order to simplify the problem, only the $N$ most significant modes are considered in the following. These are the modes with the highest modal significance 
\begin{equation}
    \mathrm{MS}_n = \frac{1}{|1 + \mathrm{j} \lambda_n|}\,,
\end{equation} 
which leads to higher contributions of the respective modes in (\ref{eq:sctFnSuperposition}). All CM data in this work is generated using our (LUH) in-house MATLAB code, which {\color{typoColor}is} evaluated in \cite{Chen2018}. 

In the following, a procedure to compare sets of far fields is proposed. It provides interpretations on why {\color{typoColor}some of these} are more suited for direction finding than others. In antenna design, this insight is used to improve the performance of {\color{typoColor}the} antenna system. 
The investigated far fields can be port far fields. If the ports are not yet defined, using Characteristic Mode far fields instead still allows the investigation of an antenna system. This approach is applicable to any multi-port antenna, including antenna arrays, but is particularly well suited for connected structures since the port setups and respective far fields are not obvious here. 
Exemplary, the procedure to identify the best suited set of CMs for direction finding using a given structure (Fig.~\ref{fig:Overview}\ref{fig:OverviewStructure}) is outlined in the following, see Fig.~\ref{fig:Overview}. It is easily modified to compare port patterns or different iterations of an antenna. All steps are detailed in the following sections. 

For a set of modes ({\color{typoColor}non-transparent} in Fig.~\ref{fig:Overview}\ref{fig:OverviewFarFields}), the CM measurement matrix $\mathbf{X}_\mathrm{CM}$ is formed. 
The DoA $k=\alpha$ is selected as reference (red dot in Fig.~\ref{fig:Overview}) and the uncertainties $u_{\alpha,\beta}$ to all test DoAs $1\leq\beta\leq K$ are determined and collected in the uncertainty vector $\mathbf{u}_\alpha$ (Fig.~\ref{fig:Overview}\ref{fig:OverviewUvec}). 
This is repeated for all possible reference DoAs $1\leq\alpha\leq K$ and the results are collected in the uncertainty matrix $\mathbf{U}$ (Fig.~\ref{fig:Overview}\ref{fig:OverviewUmat}). Next, each entry in $\mathbf{U}$ is scaled by a weighting function $w$. Subsequently, the inverse of the mean of the resulting scaled uncertainty matrix $\mathbf{U}'$ is calculated. This measure is the key performance indicator (KPI, Fig.~\ref{fig:Overview}\ref{fig:OverviewKPI}). 
To understand and improve the KPI, estimates of the incident far fields are visualized to interpret the relation between specific DoAs (Fig.~\ref{fig:Overview}\ref{fig:OverviewEstInc}). 
Finally, the KPIs of different sets of modes are compared ({\color{typoColor}transparent} in Fig.~\ref{fig:Overview}\ref{fig:OverviewKPI}) and the set with the highest resulting KPI is selected. 
The derivation of ports from the {\color{typoColor}modal} results is shown exemplary in Sec.~\ref{sec:exampleDesign}. 

\subsection{Measurement Matrix for CM: Estimated Incident Fields} \label{sec:measurementMatrix}
Estimating the direction of arrival is interpreted as the attempt to reconstruct the field incident onto an antenna system. Ideally, the result is a pencil beam of infinite directivity in the direction of arrival. However, the properties of the antenna system will alter this behavior. This is comparable to its beamforming properties, which characterizes the antenna system's scattering behavior. 

Still, in analogy to the scattered field in (\ref{eq:sctFnSuperposition}), an estimate of the $\gamma$-component of the incident far field is expressed in terms of the {\color{typoColor}CM} far fields by \cite{Grundmann2021a}: 
\begin{equation} \label{eq:FincestSum}
    F_\mathrm{inc,est,\gamma',\gamma} (\mathbf{e}_\mathrm{r}', \mathbf{e}_\mathrm{r}) = \sum\limits_{n=1}^N c_{\gamma',n} (\mathbf{e}_\mathrm{r}') F_{\gamma,n}^* (\mathbf{e}_\mathrm{r}) ~,
\end{equation}
where the complex conjugate $(\cdot)^*$ transforms the outgoing wave to an incoming wave and $c_{\gamma',n} (\mathbf{e}_\mathrm{r}')$ is the incident field modal expansion coefficient. After collecting the latter in the column vector $\mathbf{c}_{\gamma'} (\mathbf{e}_{\mathrm{r},k}')$ for the $k$th plane wave, the correlation (\ref{eq:DoACorrelations}) is reformulated to \cite{Grundmann2022}: 
\begin{equation} \label{eq:rhoFromC}
    \rho_{\alpha,\beta} = \frac{\mathbf{c}_{\gamma'} ^\mathrm{H} (\mathbf{e}_{\mathrm{r},\alpha}') \mathbf{c}_{\gamma'} (\mathbf{e}_{\mathrm{r},\beta}')}{||\mathbf{c}_{\gamma'} (\mathbf{e}_{\mathrm{r},\alpha}')|| ||\mathbf{c}_{\gamma'} (\mathbf{e}_{\mathrm{r},\beta}')||} ~. 
\end{equation}
The derivation is given in the appendix. 
Due to the orthogonality of the CM far fields \cite{Safin2013}, this is equivalent to the correlation between the vector-valued estimated incident far fields \cite{Grundmann2022}:  
\begin{multline} \label{eq:rhoFromFinc}
    \rho_{\alpha,\beta} = \\
    \frac{ \oiint 
    \mathbf{F}^*_\mathrm{inc,est,\gamma'} (\mathbf{e}'_{\mathrm{r},\alpha}, \mathbf{e}_\mathrm{r}) \cdot
    \mathbf{F}_\mathrm{inc,est,\gamma'} (\mathbf{e}'_{\mathrm{r},\beta}, \mathbf{e}_\mathrm{r}) \mathrm{d}S}
    {\sqrt{\oiint ||\mathbf{F}_\mathrm{inc,est,\gamma'} (\mathbf{e}'_{\mathrm{r},\alpha}, \mathbf{e}_\mathrm{r})||^2 \mathrm{d}S
    \oiint
    ||\mathbf{F}_\mathrm{inc,est,\gamma'} (\mathbf{e}'_{\mathrm{r},\beta}, \mathbf{e}_\mathrm{r})||^2 \mathrm{d}S }} .
\end{multline}
This relationship is a useful tool throughout the design process: If the uncertainty $u_{\alpha,\beta}$ between two DoAs is of special interest, the corresponding estimated incident far fields are visualized and investigated, see Fig.~\ref{fig:Overview}\ref{fig:OverviewEstInc}. This provides information on why the behavior is not as {\color{typoColor}desired} and helps to identify measures to overcome the issue. 

{\color{modTextColor} To determine the uncertainty parameter $u_{\alpha,\beta}$, a simpler formulation using (\ref{eq:rhoNormalized}) and
\begin{equation} \label{eq:fCMalpha}
    \mathbf{x}_\mathrm{CM,\alpha} = 
    {\left( {\begin{array}{*{20}{c}}
  F_{\gamma',1}(\mathbf{e'}_\mathrm{r,\alpha} ) & \cdots & F_{\gamma',N}(\mathbf{e'}_\mathrm{r,\alpha} )
\end{array}} \right)^{\text{T}}} 
\end{equation} 
is chosen, where $(\cdot)^\mathrm{T}$ is the transpose. For this purpose, the additional insight gained through (\ref{eq:rhoFromFinc}) is given up to ensure comparability to port far fields, since this formulation is not affected by the proportionality constant between $c_{\gamma',n}$ and $F_{\gamma',n}$.}
If, instead of the estimated incident far fields, the scattered far fields (\ref{eq:sctFnSuperposition}) are used, the correlation becomes \cite{Safin2013}: 
\begin{equation} \label{eq:rhoRealized}
    \rho_\mathrm{realized,\alpha,\beta} = 
    \frac{\mathbf{a}_{\gamma'} ^\mathrm{H} (\mathbf{e}_{\mathrm{r},\alpha}') \mathbf{a}_{\gamma'} (\mathbf{e}_{\mathrm{r},\beta}')}{||\mathbf{a}_{\gamma'} (\mathbf{e}_{\mathrm{r},\alpha}')|| ||\mathbf{a}_{\gamma'} (\mathbf{e}_{\mathrm{r},\beta}')||} ~,
\end{equation}
where the column vector $\mathbf{a}_{\gamma'}  (\mathbf{e}_{\mathrm{r},k}')$ contains the modal weighting coefficients $a_{\gamma',n} (\mathbf{e}_\mathrm{r}')$. 
These depend on the modal significance and, thereby, include the resonance behavior of the CM, see (\ref{eq:cnFromT}) in the appendix. Therefore, they tend to zero for non-significant modes, {\color{typoColor}in contrast} to the incident field modal expansion coefficients $c_{\gamma',n} (\mathbf{e}_\mathrm{r}')$. 

The influence of the modal significance on a CM far field may be viewed in analogy to the influence of the port mismatch on an antenna pattern. Therefore, the utilization of (\ref{eq:fCMalpha}) in $\mathbf{X}_\mathrm{CM}$ is treated equivalently to the utilization of the directivity or gain in $\mathbf{X}_\mathrm{Ant}$, while 
\begin{equation} \label{eq:aOverF}
    \mathbf{x}_\mathrm{CM,realized,\alpha} = 
    \left( {\begin{array}{*{20}{c}}
  \frac{F_{\gamma',1}(\mathbf{e'}_\mathrm{r,\alpha} ) }{1+\mathrm{j}\lambda_1}
  & \cdots & 
  \frac{F_{\gamma',N}(\mathbf{e'}_\mathrm{r,\alpha} ) }{1+\mathrm{j}\lambda_N}
\end{array}} \right)^{\text{T}} \,,
\end{equation}
is treated equivalently to the realized gain. The entries in (\ref{eq:aOverF}) are proportional to the modal weighting coefficients $a_{\gamma',n} (\mathbf{e}_\mathrm{r}')$, which motivates the index of $\rho_\mathrm{realized,\alpha,\beta}$ in (\ref{eq:rhoRealized}). Consequently, all parameters derived from this are assigned with a similar index. 

Depending on the stage in the design process, either (\ref{eq:fCMalpha}) or (\ref{eq:aOverF}) can be better suited. 
If the exact shape and size of the antenna structure are still to be defined, the directivity type (\ref{eq:fCMalpha}) is to be preferred. 
This way, modes of interest are identified. If required, the antenna structure is modified afterwards to shift the resonant frequency of the modes into the frequency band of interest. 
On the other hand, if only limited modifications of the antenna structure are allowed, like e.\,g. an aircraft hull, (\ref{eq:aOverF}) provides a more realistic picture of the expected behavior. This way, non-resonant modes are discarded automatically. 

Throughout this work, the entries in $\mathbf{X}_\mathrm{CM}$ are calculated from an interpolation of the discretized modal far fields. Consequently, all calculations are carried out as post-processing, which only requires the modal far fields and the eigenvalues. 

\subsection{The Uncertainty Matrix} \label{sec:uncertaintyMatrix}
Since the measurement matrix for Characteristic Modes is now defined, the uncertainty parameter (\ref{eq:rhoNormalized}) is investigated next. The uncertainty $u_{\alpha,\beta}$ between the measurement vector for a reference DoA $\alpha$, indicated by the red dot in Fig.~\ref{fig:Overview}, and the measurement vectors for all test DoAs $\beta$ is calculated and stored in the uncertainty column vector $\mathbf{u}_\alpha$. This is visualized in a spherical plot in Fig.~\ref{fig:Overview}\ref{fig:OverviewUvec}, that depicts the uncertainty for each test DoA. 

For an ideal direction finding system, the uncertainty is zero for all combinations of $\alpha$ and $\beta$, except for identical DoAs $\alpha=\beta$.  In practice however, measurement vectors of neighboring DoAs are correlated, which increases the uncertainty. A steeper decrease of the uncertainty w.\,r.\,t. the angular distance is equivalent to an improved angular resolution \cite{Grundmann2022}. 

Furthermore, a monotonous decrease of the uncertainty is desired. This way, it is ensured that an increase in the noise level leads to a steady increase of the estimation error. Otherwise, in the extreme case of an ambiguity, small errors of the measurement lead to large errors for the estimation. 
In the display of $\mathbf{u}_\alpha$, this would appear as a second maximum. 

Since the DF behavior in all selected directions is to be evaluated, each DoA in $\mathbf{X}_\mathrm{CM}$ is employed once as a reference DoA. The importance or irrelevance of specific angular regions to the direction finding application is determined by the distribution and density of the DoAs used to determine $\mathbf{X}_\mathrm{CM}$, see Fig.~\ref{fig:FarFieldMeasMat}. The resulting uncertainty column vectors $\mathbf{u}_\alpha$ are collected in the uncertainty matrix $\mathbf{U}$, depicted in Fig.~\ref{fig:Overview}\ref{fig:OverviewUmat}. {\color{newTextColor} Using (\ref{eq:fCMalpha}) in (\ref{eq:DoACorrelations}) and (\ref{eq:rhoNormalized}), its entries are given by: 
\begin{equation} \label{eq:UMatrixExplicit}
    u_{\alpha,\beta} 
    = \frac{{{{\left( {\begin{array}{*{20}{c}}
   \cdots &{{F_{\gamma ',n}^*}({{\mathbf{e}}_{{\text{r}},\alpha }'})} & \cdots  
\end{array}} \right)}}{{\left( {\begin{array}{*{20}{c}}
   \cdots &{{F_{\gamma ',n}}({{\mathbf{e}}_{{\text{r}},\beta }'})} & \cdots  
\end{array}} \right)}^{\text{T}}}}}{{\sum\limits_{n = 1}^N {{{\left\| {{F_{\gamma ',n}}({{\mathbf{e}}_{{\text{r}},\alpha }})} \right\|}^2}} \sum\limits_{n = 1}^N {{{\left\| {{F_{\gamma ',n}}({{\mathbf{e}}_{{\text{r}},\beta }})} \right\|}^2}} }} \,,
\end{equation}
with ${{F_{\gamma ',n}}({{\mathbf{e}}_{{\text{r}},\alpha }'})}$ being the $\gamma'$-component of the $n$-th Characteristic Mode far field in $\mathbf{e}_\mathrm{r,\alpha}'$ direction. }

\begin{figure}
    \color{newTextColor}
    \centering
    \includegraphics[width=0.8\columnwidth]{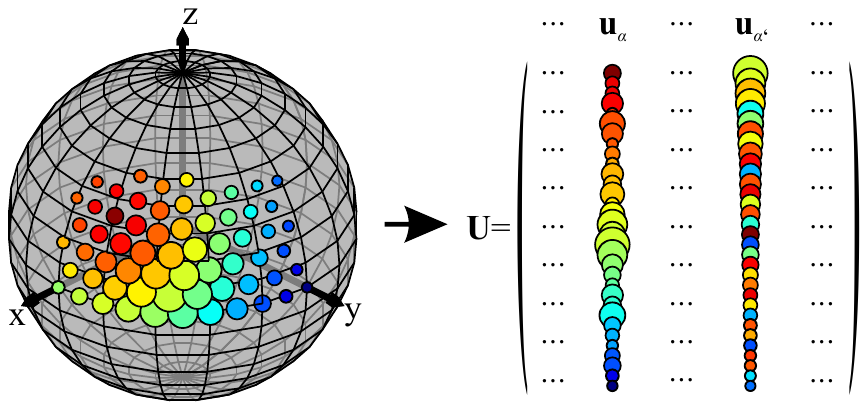}
    \caption{Illustration of the sorting of the entries in the uncertainty matrix $\mathbf{U}$ w.\,r.\,t. the great circle distance between the DoAs. The order in column $\alpha$ is indicated by the color of the marker, the order in column $\alpha'$ by its size. }
    \label{fig:SortingInU}
\end{figure}

For a visual inspection of $\mathbf{U}$, its entries must be sorted. In each column of $\mathbf{U}$, the entries are sorted w.\,r.\,t. the great circle distance between the reference DoA $\alpha$ and the test DoA $\beta$, in ascending order. The great circle distance between two DoAs $(\Theta_\alpha,\Phi_\alpha)$ and $(\Theta_\beta,\Phi_\beta)$ on a unit sphere is \cite{Grundmann2022}:  
\begin{equation} \label{eq:greatCircleDistance}
\begin{split}
    \delta_\mathrm{gc,\alpha,\beta} = \arccos &\big( \cos(\Theta_\alpha) \cos(\Theta_\beta) \\
    &+ \sin(\Theta_\alpha) \sin(\Theta_\beta) \cos(\Phi_\beta - \Phi_\alpha) \big) \,.
\end{split}
\end{equation}
The reference DoAs, and thereby the columns in $\mathbf{U}$, are sorted  w.\,r.\,t. their great circle distance to the north pole $\Theta=0^\circ$, $\Phi=0^\circ$, in increasing order{\color{newTextColor}, see Fig.~\ref{fig:SortingInU}}. 

The interpretation derived for $\mathbf{u}_\alpha$ is generalized and transferred to $\mathbf{U}$. A steep decline after the first maximum, w.\,r.\,t. an increasing test DoA index, and the absence of additional maxima are favorable properties, see Fig.~\ref{fig:Overview}\ref{fig:OverviewUmat}. 
Due to the sorting in $\mathbf{U}$, the uncertainty for the first test DoA is inversely proportional to the square root of the total power received for the respective reference DoA. Therefore, the maximum of the uncertainty matrix is not necessarily equal for all reference DoAs. 

\subsection{A Key Performance Indicator to Compare Sets of Ports and Modes} \label{sec:KPI}
In order to be able to compare large quantities of different sets of {\color{typoColor}Characteristic Modes}, antenna ports or even antennas, it is not suitable to rely on a visual inspection of the uncertainty matrices. Instead, a single parameter is required to summarize the performance. This key performance indicator (KPI) reflects a compromise between the need for a low correlation of the measurement vectors in different angular regions and a constant and high received power. If required, the weights of these components can be varied depending on the application. The derivation of the KPI is detailed in the following. 

Since ambiguities shall be avoided, a higher weight is applied to test DoAs that are further away from the reference DoA than to neighboring DoAs. Therefore, a linear weighting function is introduced \cite{Grundmann2022}:
\begin{equation} \label{eq:weightingFunction}
    w(\delta_\mathrm{gc,\alpha,\beta}) = \frac{\delta_\mathrm{gc,\alpha,\beta}}{\pi} \,.
\end{equation}
{\color{newTextColor}This yields zero for identical test and reference DoAs and unit weight for the maximum great circle distance $\delta_\mathrm{gc,\alpha,\beta}=180^\circ$, see~(\ref{eq:greatCircleDistance}).}
If higher penalties for {\color{typoColor}specific} DoAs are desired, this is replaced by a different function. Eq.~\ref{eq:weightingFunction} is used to determine the entries of a matrix $\mathbf{W}$, of which the Hadamard, or element-wise, product with $\mathbf{U}$ is calculated to obtain the weighted uncertainty matrix \cite{Grundmann2022}: 
\begin{equation} \label{eq:Cprime}
    \mathbf{U}' = \mathbf{W}\, \circ\, \mathbf{U} \,.
\end{equation}

Now, we choose the inverse of the mean of the elements of $\mathbf{U}'$ as the key performance indicator: 
\begin{equation} \label{eq:KPI}
    \mathrm{KPI} 
    = \frac{1}{\mathrm{mean}(\mathbf{U}')}
    = \left( \frac{1}{K^2} \sum\limits_{\alpha = 1}^K \sum\limits_{\beta = 1}^K |u_{\alpha,\beta}|\, w(\delta_\mathrm{gc,\alpha,\beta}) \right)^{-1} ,
\end{equation}
so a higher KPI indicates that a setup is better suited for direction finding, see Fig.~\ref{fig:Overview}\ref{fig:OverviewKPI}. {\color{newTextColor}In contrast to the parameter proposed in \cite{Grundmann2022}, the KPI is invariant to the angular resolution of the DoAs.}

{\color{newTextColor}However, the aim of the KPI is not to provide an accurate prediction of the root mean square error (RMSE) of an actual DoA estimation. The RMSE depends on the utilized algorithm and how it is affected by noise. If a closer resemblance with the RMSE of a specific algorithm is desired, (\ref{eq:weightingFunction}) can be modified to be a function of $u_{\alpha,\beta}$. The parameters of this function need to be tuned to match the behavior of the investigated algorithm and signal to noise ratio (SNR). Since this work aims to provide a deterministic approach that only depends on the antenna properties,
}{\color{modTextColor}here, the simpler definition of the KPI using (\ref{eq:KPI}) and (\ref{eq:weightingFunction}) is utilized to identify the best suited set of CM for direction finding. }

\section{Performance Evaluation Examples} \label{sec:examplesEvaluation}
\begin{figure}
\centering
\subfloat[(a)]{
\includegraphics[width=0.4\columnwidth]{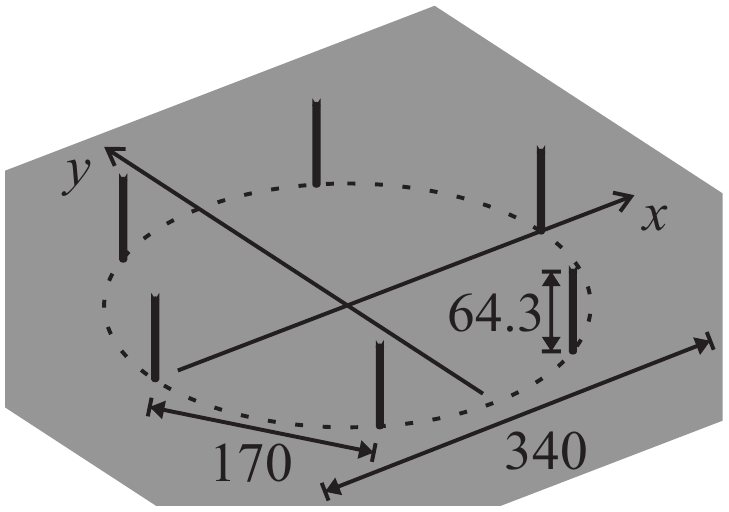}
\label{fig:CADArray.6lambda}}
\hfil
\subfloat[(b)]{
\includegraphics[width=0.4\columnwidth]{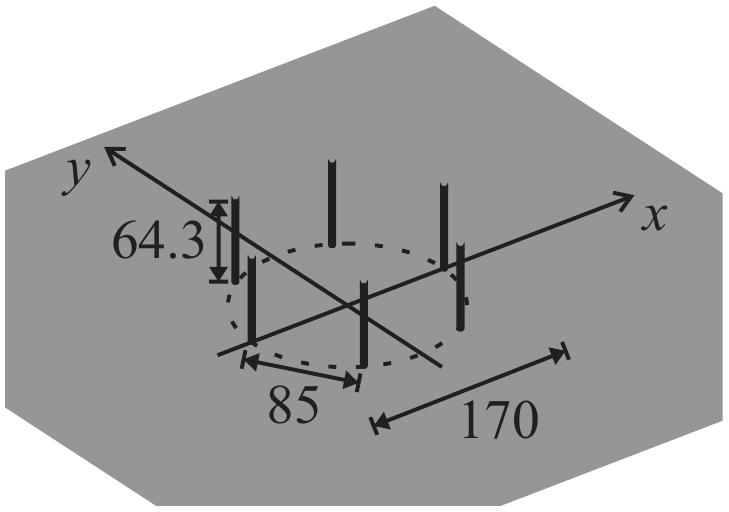}
\label{fig:CADArray.3lambda}} 
\hfil
\subfloat[(c)]{
\includegraphics[width=0.4\columnwidth]{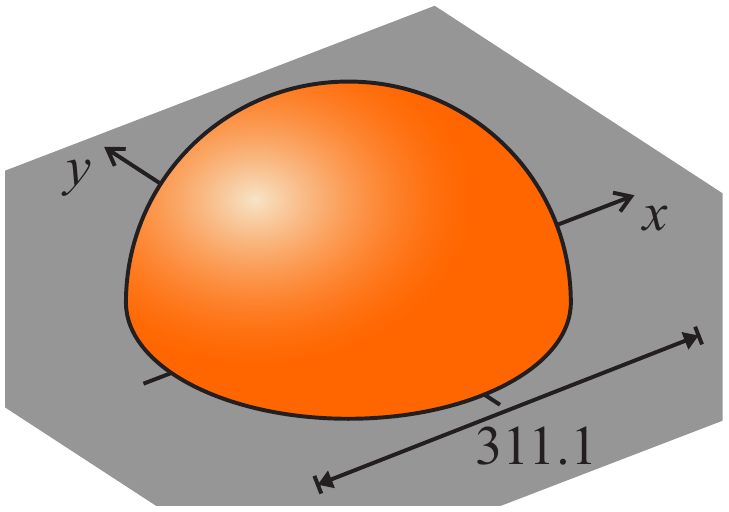}
\label{fig:CADHemisphere}} 
\hfil
\subfloat[(d)]{
\includegraphics[width=0.4\columnwidth]{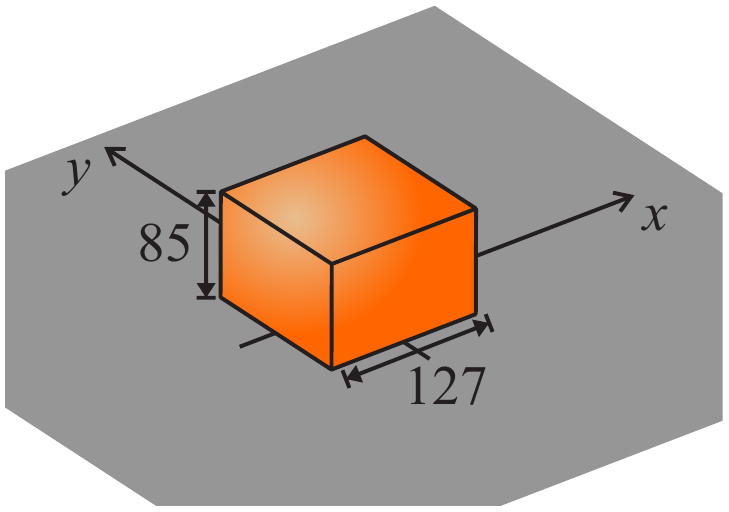}
\label{fig:CADCuboid}} 
\caption{Investigated example antenna structures consisting of PEC on an infinite ground plane. For analysis of port patterns: {\color{typoColor}uniform circular array (UCA)} of six monopole antennas with inter-element spacing of $0.6\,\lambda_\mathrm{FS}$ and $0.3\,\lambda_\mathrm{FS}$, respectively: \protect\subref{fig:CADArray.6lambda} and \protect\subref{fig:CADArray.3lambda}. For CM analysis: Hemisphere of $1.1\lambda_\mathrm{FS}$ diameter: \protect\subref{fig:CADHemisphere} and cuboid: \protect\subref{fig:CADCuboid}. 
All dimensions in are mm, depictions are to scale. } 
\label{fig:CADExamples}
\end{figure}
The focus of the investigated examples, depicted in Fig.~\ref{fig:CADExamples}, is towards aerial direction finding systems. In the following, parameters are chosen to be reasonable for this application. However, the presented method is not limited to the application nor to the parameters and is applicable to other setups. {\color{typoColor}For} Figs.~\ref{fig:CADArray.6lambda} and \ref{fig:CADArray.3lambda}, port patterns are investigated. {\color{typoColor}For} Figs.~\ref{fig:CADHemisphere} and \ref{fig:CADCuboid}, CMs are investigated with the aim of developing a set of suited antenna ports in Sec.~\ref{sec:exampleDesign}. 

Most direction finding antenna systems are not able to cover all polarizations and directions of arrival equally. {\color{newTextColor}The presented approach is applicable to designs with dual polarization. }In many occasions, however, only a single polarization and a limited angular range in azimuth and elevation need to be covered. A common scenario is the presence of a ground plate, which is often flat, electrically large and approximated by an infinite, perfectly electrically conducting (PEC) plane. In the context of aerial direction finding, this reflects the hull of the aircraft. 

Throughout this paper, the performance of the different antenna systems is therefore evaluated exemplary in the range of $45^\circ \leq \Theta \leq 90^\circ$ and $0^\circ \leq \Phi < 360^\circ$, ref. the coordinate system in Fig.~\ref{fig:introduction}a. Only the $\Theta$-polarization is considered. To obtain $\mathbf{X}_\mathrm{Ant}$ and $\mathbf{X}_\mathrm{CM}$, the far field values of the respective antenna ports and Characteristic Modes are considered in $K=250$ different directions of arrival \cite{Grundmann2022}, see Fig.~\ref{fig:FarFieldMeasMat}. These are positioned on a homogeneous grid, obtained from a recursive partitioning algorithm \cite{Gagarinov2017}. 
The investigation frequency is chosen as $1060$\,MHz. This is the center frequency between the two bands used for the Mode-S secondary radar and the ACAS \cite{ICAO2006}. {\color{newTextColor}A constant behavior of the antenna system over the relevant bandwidth is assumed in the following. For applications where this assumption does not hold, the results of the proposed procedure at multiple frequencies need to be considered w.\,r.\,t. the requirements.}

\subsection{Monopole UCA} \label{sec:monopoleUCA}
In order to verify the properties and behavior of the proposed procedure and compare it to established methods, a simple and common example is investigated first. This is chosen to be a uniform circular array (UCA) of six monopole antennas, positioned on an infinite PEC plane. The $z$-directed monopole antennas have a length of $64.3$\,mm each, so they are matched to the port impedance of $50\,\Omega$ at the investigated frequency. The ports are positioned at the foot points of the monopoles and the far fields are determined from a finite differences time domain (FDTD) simulation of the complete array in EmpireXPU \cite{empirexpu}. Notice that the coupling between the antennas is included, while the mismatch and therefore the influence of the active impedance on the patterns is neglected here. 

\begin{figure}
\centering
\subfloat[(a)]{
    \includegraphics[valign=c]{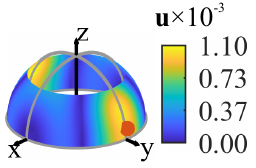} 
\label{fig:uncert6MonopoleArray.6lambda}} 
\hfil
\subfloat[(b)]{
    \includegraphics[valign=c]{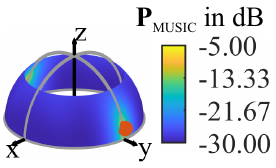} 
\label{fig:MUSICspectrum6MonopoleArray.6lambda}} 
\hfil
\subfloat[(c)]{
    \includegraphics[valign=c]{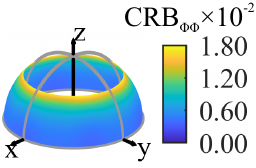} 
\label{fig:CRB6MonopoleArray.6lambda}} 
\caption{Results of different evaluation procedures for a UCA of six quater-wavelength monopoles with inter-element distance $0.6\lambda_\mathrm{FS}$. Uncertainty: \protect\subref{fig:uncert6MonopoleArray.6lambda} and MUSIC spectrum: \protect\subref{fig:MUSICspectrum6MonopoleArray.6lambda}, calculated for a $\Theta$-polarized wave incident from $(80^\circ,90^\circ)$, indicated by the red dot.
CRB for estimating $\Phi$ in $\mathrm{deg}/100$: \protect\subref{fig:CRB6MonopoleArray.6lambda}. The SNR is 0\,dB and a single sample is investigated. 
} \label{fig:Comparison6MonopoleArray}
\end{figure}
Firstly, the UCA from Fig.~\ref{fig:CADArray.6lambda} with an inter-element distance of $0.6\lambda_\mathrm{FS}$ is investigated, with $\lambda_\mathrm{FS}$ being the free space wavelength. The uncertainty vector $\mathbf{u}_\alpha$ for a plane wave incident from $(\Theta=80^\circ,\Phi=90^\circ)$ is displayed in Fig.~\ref{fig:uncert6MonopoleArray.6lambda}. The discrete values {\color{typoColor}obtained} at the different DoAs are interpolated in a surface plot. On the opposite side of the reference DoA, a second maximum is visible, indicating an ambiguity. Since the described array setup has significant grating lobes, this observation matches the expectations. 

{\color{modTextColor}
Secondly, the result of the MUSIC algorithm \cite{Schmidt1986}, the so called MUSIC spectrum, is depicted in Fig.~\ref{fig:MUSICspectrum6MonopoleArray.6lambda}. As is concluded from the second maximum at ${\Phi=270^\circ}$, the ambiguity is also detected by the MUSIC algorithm.
However, the dynamic range of the MUSIC spectrum is much higher than that of the uncertainty vector, in particular for low noise levels. This behavior is evident since the aim of the algorithm is to reduce the influence of undesired side lobes. For antenna development however, the aim is to magnify these influences in the evaluation procedure, in order to be able to detect and eliminate them with a better suited antenna design. Thereby, the MUSIC algorithm makes visual inspection of the results more difficult and time consuming Monte Carlo simulations are required to obtain accurate results. 
Moreover, the intuitive visualization of the incident fields (Fig.~\ref{fig:Overview}\ref{fig:OverviewEstInc}) is lost in this case.  

Finally, a depiction of the CRB \cite{Weiss1991} is given in Fig.~\ref{fig:CRB6MonopoleArray.6lambda}. Since the CRB does not inherently cover ambiguities \cite{Gazzah2006, Ma2022}, none are observable here. This is evident since the CRB for a specific DoA is not affected by the far fields in other directions.}

\begin{figure}
\centering
\subfloat[(a)]{\noindent
    \begin{tikzpicture}[baseline]
    \node[right, inner sep=0] (uMat) at (0,0) {
    \includegraphics[valign=c]{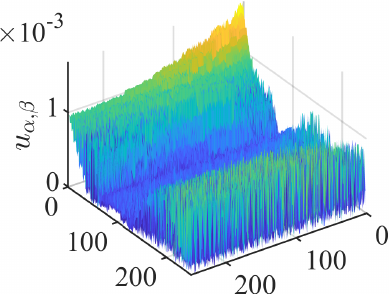} };
    \node (tstDoA) at (0.5, -1.4) {\footnotesize \begin{tabular}{l} Test\\ DoA\end{tabular}};
    \node (refDoA) at (3.5, -1.4) {\footnotesize Ref. DoA};
    \end{tikzpicture}%
\label{fig:C6MonopoleArray.6lambda}}%
\hfil
\subfloat[(b)]{\noindent
    \begin{tikzpicture}[baseline]
    \node[right, inner sep=0] (uMat) at (0,0) {
    \includegraphics[valign=c]{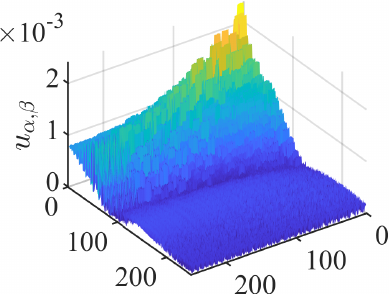} };
    \node (tstDoA) at (0.5, -1.4) {\footnotesize \begin{tabular}{l} Test\\ DoA\end{tabular}};
    \node (refDoA) at (3.5, -1.4) {\footnotesize Ref. DoA};
    \end{tikzpicture}%
\label{fig:C6MonopoleArray.3lambda}}%
\caption{Uncertainty matrices $\mathbf{U}$ of a UCA of six quarter-wavelength monopoles. Inter-element distances are $0.6\lambda_\mathrm{FS}$ \protect\subref{fig:C6MonopoleArray.6lambda} and $0.3\lambda_\mathrm{FS}$ \protect\subref{fig:C6MonopoleArray.3lambda}. } \label{fig:C6MonopoleArray}
\end{figure}
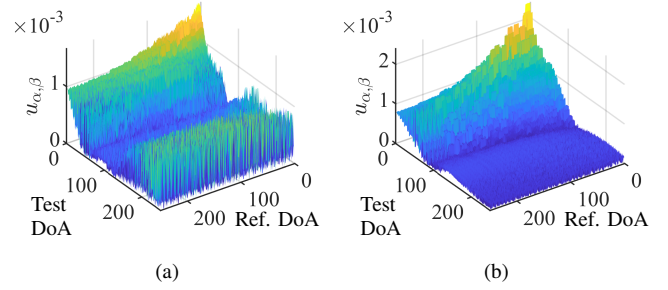
{\color{modTextColor}Since the proposed procedure provides more insight into the behavior of the antenna system than the MUSIC algorithm or the CRB, it is used to investigate the observed ambiguous behavior for all DoAs. Therefore, the uncertainty matrix of the investigated setup is determined and shown in Fig.~\ref{fig:C6MonopoleArray.6lambda}.
Here, a higher index of the test DoA corresponds to an increased great circle distance to the respective reference DoA. With increasing distance, at first a steep decrease of $u_{\alpha,\beta}$ is observed, followed by a second maximum for all reference DoAs. This confirms the expected ambiguity. }

Since grating lobes disappear for inter-element distances smaller than $0.5\lambda_\mathrm{FS}$, the ambiguity is expected to disappear as well in this case. This is reflected by Fig.~\ref{fig:C6MonopoleArray.3lambda}, which depicts the uncertainty matrix of a similar UCA with an inter-element spacing of $0.3\lambda_\mathrm{FS}$, see Fig.~\ref{fig:CADArray.3lambda}. In Fig.~\ref{fig:C6MonopoleArray.3lambda}, no second maximum is observed. However, the decrease of the uncertainty parameter is more gradual. This is attributed to a reduction of the maximum directivity due to the reduced aperture size. 

For both UCAs, the $u_{\alpha,\beta}$ decreases with increasing reference DoA indices, see Fig.~\ref{fig:C6MonopoleArray}. The reason for this is the increasing directivity as the elevation angle approaches the ground plane, which decreases the uncertainty parameter according to (\ref{eq:rhoNormalized}). 

While the proposed method is suitable to evaluate the performance of antenna arrays, it does not develop its full potential as a design assistance for simple setups like this. Therefore, the more complex case of a single radiating antenna structure is investigated next. 

\subsection{Hemispherical Shell} \label{sec:hemisphere}
\begin{figure}
\centering
\subfloat[Mode 1]{
\includegraphics[]{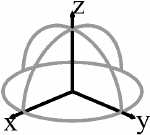}
\label{fig:hemisphereMode1}}
\hfil
\subfloat[Mode 2]{
\includegraphics[]{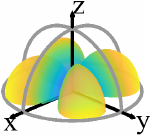}
\label{fig:hemisphereMode2}}
\hfil
\subfloat[Mode 3]{
\includegraphics[]{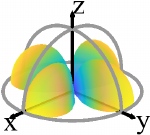}
\label{fig:hemisphereMode3}}
\hfil
\subfloat[Mode 4]{
\includegraphics[]{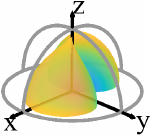}
\label{fig:hemisphereMode4}}
\hfil
\subfloat[Mode 5]{
\includegraphics[]{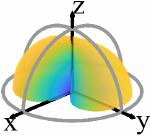}
\label{fig:hemisphereMode5}}
\hfil
\subfloat[Mode 6]{
\includegraphics[]{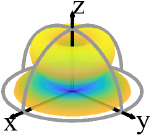}
\label{fig:hemisphereMode6}}
\hfil
\subfloat[Mode 7]{
\includegraphics[]{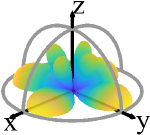}
\label{fig:hemisphereMode7}}
\hfil
\subfloat[Mode 8]{
\includegraphics[]{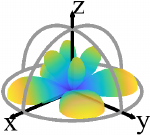}
\label{fig:hemisphereMode8}}
\hfil
\subfloat[Mode 9]{
\includegraphics[]{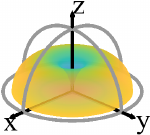}
\label{fig:hemisphereMode9}}
\hfil
\subfloat[]{
\includegraphics[]{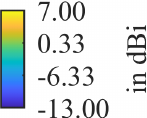}
\label{fig:hemisphereModeCbar}}
\caption{$\Theta$-component of the far field directivity for the nine most significant modes of a hemispherical shell with diameter $1.1\lambda_\mathrm{FS}$ on an infinite ground plane. The $\Theta$-component of the first mode is approximately zero in all directions. } \label{fig:ModesHemisphere}
\end{figure}
The largest antenna structure fitting inside a hemispherical radome like {\color{typoColor}the one} in Fig.~\ref{fig:introduction}a is a PEC hemisphere. Therefore, the hemispherical shell with a diameter of $1.1\lambda_\mathrm{FS}$ from Fig.~\ref{fig:CADHemisphere} is investigated next. It is positioned on an infinite ground plane. The $\Theta$-components of its nine most significant Characteristic Mode far fields are depicted in Fig.~\ref{fig:ModesHemisphere}. The $\Theta$-component of the first mode is zero. Still, it is included in the following evaluations to demonstrate that it is not selected as a suited mode. 

\begin{figure}
\centering
\subfloat[]{
\includegraphics[valign=c]{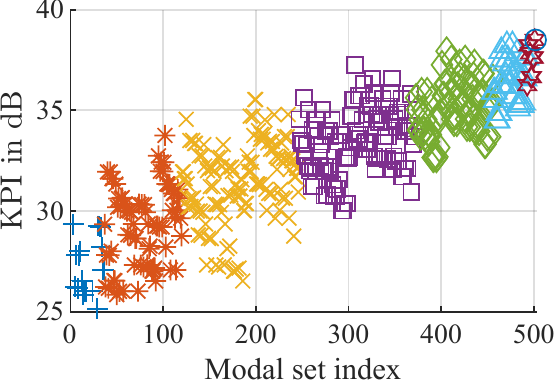}
\label{fig:hemisphereOpt}}
\hfil
\subfloat[]{
\includegraphics[valign=c]{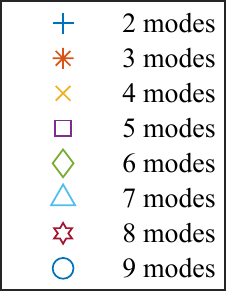}
\label{fig:hemisphereOptLeg}}
\caption{Comparison of the KPI for different sets of modes in the hemispherical shell example. } \label{fig:OptHemisphere}
\end{figure}
\begin{table}
\renewcommand{\arraystretch}{1.3}
\caption{Best performing sets of modes\\ for the hemispherical shell example}
\label{tab:hemisphereModalSets}
\centering
\begin{tabular}{|c|c|c|}
    \hline
    \bfseries No. of Modes & \bfseries Best modal set & \bfseries KPI in dB\\
    \hline\hline
2 & $2, 3$ & 29.34\\
\hline
3 & $4, 5, 9$ & 33.76\\
\hline
4 & $[2, 3], 4, 5, 9$ & 35.54\\
\hline
5 & $2, 3, 4, 5, 9$ & 37.26\\
\hline
6 & $2, 3, 4, 5, 6, 9$ & 37.94\\
\hline
7 & $2, 3, 4, 5, 6, [7, 8], 9$ & 38.22\\
\hline
8 & $2, 3, 4, 5, 6, 7, 8, 9$ & 38.50\\
\hline
9 & $1, 2, 3, 4, 5, 6, 7, 8, 9$ & 38.50\\
\hline
\end{tabular}
\end{table}
For all possible combinations of modes, the resulting KPI is displayed in Fig.~\ref{fig:OptHemisphere}. The results are grouped by the number of utilized modes, indicated by different colors and markers. The set with the highest value for each color is the best performing set for the selected number of modes. The respective modes and the exact KPIs are given in Table~\ref{tab:hemisphereModalSets}. Here, square brackets indicate that either of the modes may be selected, which is due to a degeneracy of the modes. 

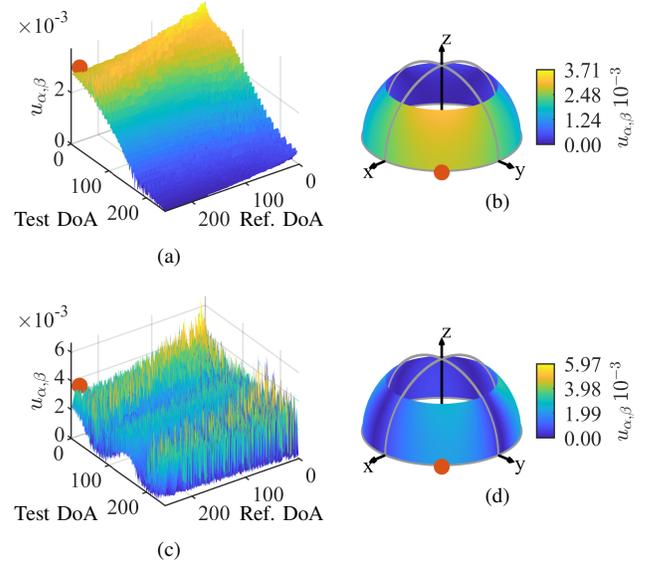
\begin{figure}
\centering
\subfloat[(a)]{
    \begin{tikzpicture}[baseline]
    \node[right, inner sep=0] (uMat) at (0,0) {
    \includegraphics[valign=c]{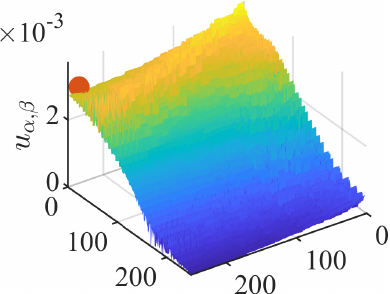} };
    \node (tstDoA) at (0.5, -1.4) {\footnotesize Test DoA};
    \node (refDoA) at (3.5, -1.4) {\footnotesize Ref. DoA};
    \end{tikzpicture}
\label{fig:hemisphereC459}}
\hfil
\subfloat[(b)]{
\includegraphics[valign=c] {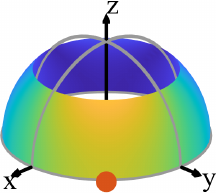}
\includegraphics[valign=c]{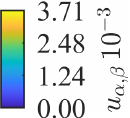}
\label{fig:hemisphereC459col235}}
\hfil
\subfloat[(c)]{
    \begin{tikzpicture}[baseline]
    \node[right, inner sep=0] (uMat) at (0,0) {
    \includegraphics[valign=c]{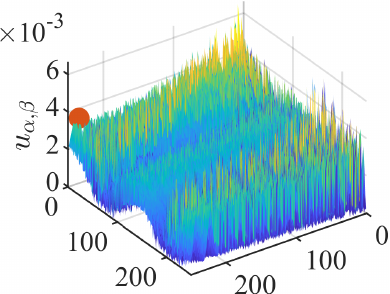} };
    \node (tstDoA) at (0.5, -1.4) {\footnotesize Test DoA};
    \node (refDoA) at (3.5, -1.4) {\footnotesize Ref. DoA};
    \end{tikzpicture}
\label{fig:hemisphereC234}}
\hfil
\subfloat[(d)]{
\includegraphics[valign=c] {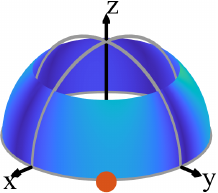}
\includegraphics[valign=c]{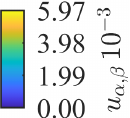}
\label{fig:hemisphereC234col235}}
\caption{Results for a hemispherical shell with $1.1\lambda_\mathrm{FS}$ diameter on an infinite ground plane. Uncertainty matrices $\mathbf{U}$ are depicted in \protect\subref{fig:hemisphereC459} and \protect\subref{fig:hemisphereC234}, while the uncertainty vectors $\mathbf{u}_\alpha$ for the reference DoA ($\Theta=90^\circ,\Phi=45^\circ$) are shown in \protect\subref{fig:hemisphereC459col235} and \protect\subref{fig:hemisphereC234col235}. 
The selected modes are: $\lbrace 4, 5, 9\rbrace$ in \protect\subref{fig:hemisphereC459} and \protect\subref{fig:hemisphereC459col235}; as well as $\lbrace 2,3,4\rbrace$ in \protect\subref{fig:hemisphereC234} and \protect\subref{fig:hemisphereC234col235}.} \label{fig:hemisphereCs}
\end{figure}
Table~\ref{tab:hemisphereModalSets} reveals that the maximum achievable performance improves with an increasing number of modes. The only exception is the addition of mode~1 in the last step, which confirms that it is not suited for direction finding {\color{newTextColor}since its $\Theta$-component is zero}. However, Fig.~\ref{fig:OptHemisphere} shows that for a given number of modes, the performance varies heavily between different sets of modes. First, we investigate the set of modes $\lbrace 4, 5, 9\rbrace$, which is the best performing set of three modes, by using the uncertainty matrix depicted in Fig.~\ref{fig:hemisphereC459}. It is observed that there is no ambiguity, i.\,e. the uncertainty declines monotonically with increasing test DoA index. This is also reflected in Fig.~\ref{fig:hemisphereC459col235}, which depicts the uncertainty parameter for the reference DoA ($\Theta=90^\circ,\Phi=45^\circ$), which is marked by the red dot. 

Now, the best performing set of three modes without mode~9 is investigated, which consists of the modes $\lbrace 2, 3, 4\rbrace$. By looking at the corresponding uncertainty matrix depicted in Fig.~\ref{fig:hemisphereC234}, a three-fold ambiguity of varying intensity is identified for all DoAs. This is confirmed by investigating the columns of $\mathbf{U}$, like the one for the reference DoA ($90^\circ,45^\circ$), depicted in Fig.~\ref{fig:hemisphereC234col235}. Here, the two additional maxima {\color{typoColor}representing the ambiguous behavior} are visible at $\Phi\approx-45^\circ$ and $\Phi\approx 135^\circ$, to the left- and right-hand sides of the plot. It is evident that this set of modes is not suited for direction finding in the specified angular region. However, if mode~9 had not been included in the optimization, this would have been the best suited set, based on the KPI. Preventing a false selection is possible since the proposed method provides a way to investigate a set of modes beyond this scalar performance metric. 

\begin{figure}
\centering
\subfloat[(a)]{
    \begin{tikzpicture}[baseline]
    \node[right, inner sep=0] (uMat) at (0,0) {
    \includegraphics[valign=c]{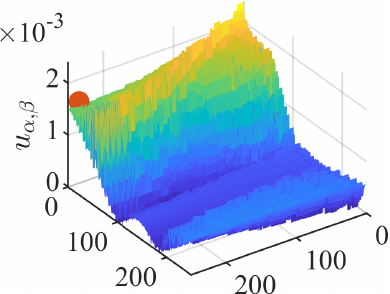} };
    \node (tstDoA) at (0.5, -1.4) {\footnotesize Test DoA};
    \node (refDoA) at (3.5, -1.4) {\footnotesize Ref. DoA};
    \end{tikzpicture}
\label{fig:hemisphereC23459}}
\hfil
\subfloat[(b)]{
\includegraphics[valign=c]{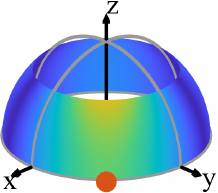}
\includegraphics[valign=c]{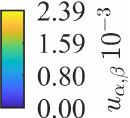}
\label{fig:hemisphereC23459col235}}
\hfil
\subfloat[(c)]{
\includegraphics[valign=c]{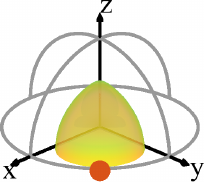}
\label{fig:hemisphereC23459Einc235}}
\hfil
\subfloat[(d)]{
\includegraphics[valign=c]{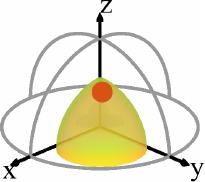}
\label{fig:hemisphereC23459Einc7}}
\hfil
\subfloat[(e)]{
\includegraphics[valign=c]{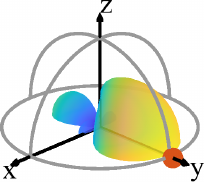}
\label{fig:hemisphereC23459Einc240}}
\hfil
\subfloat[]{
\includegraphics[valign=c]{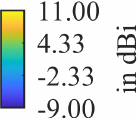} }
\caption{Results for a hemispherical shell with $1.1\lambda_\mathrm{FS}$ diameter on an infinite ground plane, using modes $\lbrace 2, 3, 4, 5, 9\rbrace$. \protect\subref{fig:hemisphereC23459}: uncertainty matrix $\mathbf{U}$, \protect\subref{fig:hemisphereC23459col235}: uncertainty vector $\mathbf{u}_\alpha$ for the reference DoA ($\Theta=90^\circ,\Phi=45^\circ$), \protect\subref{fig:hemisphereC23459Einc235}: estimated incident field for that DoA; \protect\subref{fig:hemisphereC23459Einc7} and \protect\subref{fig:hemisphereC23459Einc240}: estimated incident fields for $(45^\circ,45^\circ)$ and $(0^\circ,90^\circ)$, respectively. The red markers show the currently investigated reference DoA. } \label{fig:hemisphereCs5modes}
\end{figure}

Finally, the results for the set containing modes $\lbrace 4, 5, 9\rbrace$ are compared to those for the best performing set of five modes, being $\lbrace 2, 3, 4, 5, 9\rbrace$, depicted in Fig.~\ref{fig:hemisphereCs5modes}. From Fig.~\ref{fig:hemisphereC23459}, we observe that the slope of the uncertainty matrix is steeper than in Fig.~\ref{fig:hemisphereC459}, improving the direction finding performance. This is {\color{typoColor}confirmed by} Fig.~\ref{fig:hemisphereC23459col235}, where the uncertainty vector $\mathbf{u}_\alpha$ for the reference DoA $(90^\circ, 45^\circ)$ is displayed. 
However, while the uncertainty decreases monotonically in azimuthal direction, it increases slightly for elevation angles towards the $z$-axis. In this region, the amplitude of the far fields is lower than in the $x$-$y$-plane, increasing the uncertainty. 
By investigating the estimated incident fields in Figs.~\ref{fig:hemisphereC23459Einc235} - \ref{fig:hemisphereC23459Einc240}, it is found that the correlation between the measurement vectors also contributes to this effect: There is little difference between the estimated incident fields for DoAs with the same azimuth but different elevations, compare Figs.~\ref{fig:hemisphereC23459Einc235} and \ref{fig:hemisphereC23459Einc7}. On the other hand, a change of the azimuth angle by the same distance yields a significantly different estimated incident field, compare Figs.~\ref{fig:hemisphereC23459Einc235} and \ref{fig:hemisphereC23459Einc240}.
Concluding, the investigated set of modes has a limited ability to estimate the elevation of an incoming signal. 

In practical antenna design, the number of available coherent receivers is usually limited, which in turn limits the number of ports. From the results explained above, it is concluded that three or five ports are a good compromise between hardware effort and direction finding performance. Also, candidate modes for an implementation are identified. 

\section{Antenna System Design Example} \label{sec:exampleDesign}
While a hemispherical shell is an easily-defined example, it is difficult to manufacture, compared to polygonal structures. For simplicity, a cuboid antenna shape with a square footprint is chosen as the basis for the development and manufacturing of a demonstration antenna system. Like the other example structures, the cuboid is positioned on an infinite ground plane, but no further restrictions are imposed on the geometry. 
It is assumed that three coherent receivers are available. Therefore, a Multi-Mode Multi-Port Antenna (M$^3$PA), also known as Multi-Mode Antenna (MMA) is developed. It will provide three mutually uncorrelated antenna ports that aim for far field patterns similar to the selected Characteristic Modes. 

\subsection{Design and Description}
After the general shape of the structure is defined, suitable dimensions need to be determined. For the hemispherical shell from Sec.~\ref{sec:hemisphere}, the combination of the two magnetic dipole modes 4 and 5 with the electric monopole mode 9, see Fig.~\ref{fig:ModesHemisphere}, is selected. Since the electromagnetic behavior of the cuboid is related to that of the hemisphere, it is reasonable to use these three modes as a starting point. Consequently, the dimensions of the cuboid are chosen so these three modes are significant. 
The resulting dimensions are depicted in Fig.~\ref{fig:CADCuboid}. 
Since the cuboid is smaller, the diameter of the circumscribing radome is reduced by twenty percent, compared to the hemispherical shell. 

\begin{figure}
\centering
\subfloat[Mode 1]{
\includegraphics[]{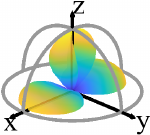}
\label{fig:cuboidMode1}}
\hfil
\subfloat[Mode 2]{
\includegraphics[]{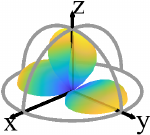}
\label{fig:cuboidMode2}}
\hfil
\subfloat[Mode 3]{
\includegraphics[]{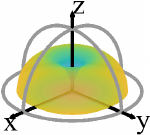}
\label{fig:cuboidMode3}}
\hfil
\subfloat[Mode 4]{
\includegraphics[]{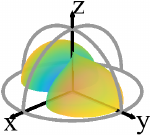}
\label{fig:cuboidMode4}}
\hfil
\subfloat[Mode 5]{
\includegraphics[]{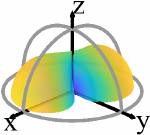}
\label{fig:cuboidMode5}}
\hfil
\subfloat[Mode 6]{
\includegraphics[]{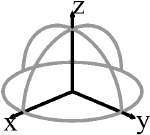}
\label{fig:cuboidMode6}}
\hfil
\subfloat[Mode 7]{
\includegraphics[]{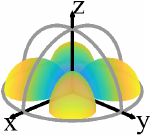}
\label{fig:cuboidMode7}}
\hfil
\subfloat[Mode 8]{
\includegraphics[]{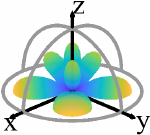}
\label{fig:cuboidMode8}}
\hfil
\subfloat[Mode 9]{
\includegraphics[]{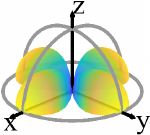}
\label{fig:cuboidMode9}}
\hfil
\subfloat[]{
\includegraphics[]{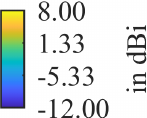}
\label{fig:cuboidModesCbar}}
\caption{$\Theta$-component of the far field directivity at 1060\,MHz for the nine most significant modes of a square cuboid with edge length 127\,mm and height 85\,mm on an infinite ground plane. The $\Theta$-component of mode 6 is approximately zero in all directions.} \label{fig:cuboidModes}
\end{figure}
The $\Theta$-components of the Characteristic Mode far fields of the nine most significant modes of the cuboid are depicted in Fig.~\ref{fig:cuboidModes}. Here, the electric monopole mode is number 3, while the magnetic dipole modes are numbers 4 and 5. 

In order to determine whether the best suited set of modes has changed for the new structure, the proposed evaluation method is applied to the cuboid. 
Since the dimensions of the investigated structure are now fixed and the resonance behavior of the modes will not be altered significantly anymore, it is reasonable to use the realized uncertainty and KPI, based on the definition in (\ref{eq:aOverF}). This way, non-significant modes contribute less. 
The realized KPI of the set containing modes $\lbrace 3,4,5\rbrace$ is $\mathrm{KPI_{realized}} \approx 31.9$\,dB, which is the highest for any set of three modes. The realized uncertainty matrix $\mathbf{U}_\mathrm{realized}$ is qualitatively comparable to Fig.~\ref{fig:hemisphereC459} and no ambiguities are observed. 

In the following, a port setup is developed based on the selected modes, following the procedure from \cite{Peitzmeier2019}. By using group theory and symmetry, uncorrelated antenna ports are realized. In Schoenfliess notation \cite{Cornwell1997}, the symmetry group of a cuboid with square footprint on a ground plane is the $C_\mathrm{4,v}$ group. For this symmetry group, port patterns remain uncorrelated even if the investigated range of $\Theta$ does not cover the complete sphere, which is advantageous for the direction finding. 
{\color{typoColor}From symmetry theory, it is known} that any two functions belonging to two different so called irreducible representations $\Gamma^{(\kappa)}$ are uncorrelated. Since the port far fields shall be correlated to the selected modes, each pair of port and mode must belong to the same irreducible representation. 

\begin{figure}
\centering
\subfloat[(a)]{
\includegraphics[valign=c, width=0.22\columnwidth]{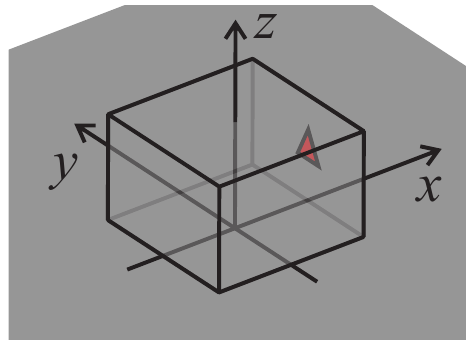}
\label{fig:cuboidCADPgen}}
\hfil
\subfloat[(b)]{
\includegraphics[valign=c, width=0.22\columnwidth]{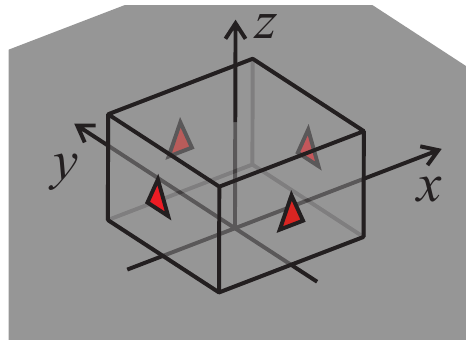}
\label{fig:cuboidCADP1}}
\hfil
\subfloat[(c)]{
\includegraphics[valign=c, width=0.22\columnwidth]{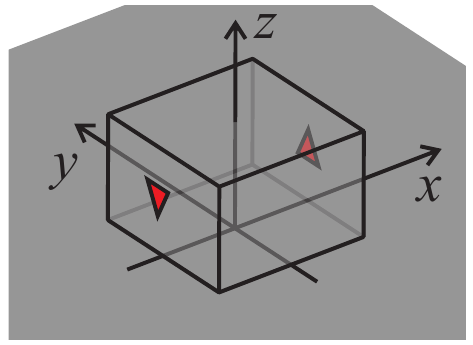}
\label{fig:cuboidCADP2}}
\hfil
\subfloat[(d)]{
\includegraphics[valign=c, width=0.22\columnwidth]{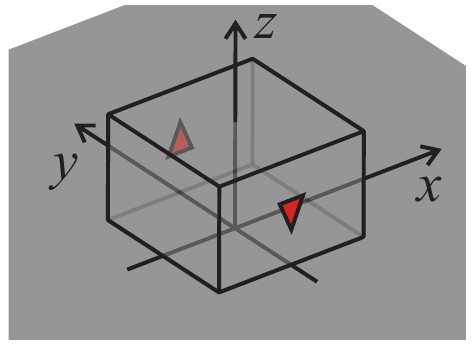}
\label{fig:cuboidCADP3}}
\hfil
\caption{Port generation on the cuboid. The $z$-directed delta gap source (red arrow) depicted in \protect\subref{fig:cuboidCADPgen} is used to generate port 1 \protect\subref{fig:cuboidCADP1} as a basis function of $\Gamma^{(1)}$, as well as port 2 \protect\subref{fig:cuboidCADP2} and port 3 \protect\subref{fig:cuboidCADP3} as mutually orthogonal basis functions of $\Gamma^{(5)}$. } \label{fig:cuboidCADPorts}
\end{figure} 
Following \cite{Peitzmeier2019}, a $z$-directed delta gap source is positioned on one of the lateral surfaces of the cuboid, see Fig.~\ref{fig:cuboidCADPgen}. In the following, these are called feed points in order to distinguish them from the uncorrelated antenna ports.
The procedure from \cite{Peitzmeier2019} is applied to generate one port from $\Gamma^{(1)}$, correlated to mode 3, and, due to degeneracy, two ports from $\Gamma^{(5)}$, correlated to modes 4 and 5. 
The data from \cite{Cornwell1997} is used for this and the resulting feed point setups are depicted in Fig.~\ref{fig:cuboidCADPorts}. Since modes 1 and 2 also have a high modal significance and belong to $\Gamma^{(5)}$, they are expected to influence the shape of the {\color{typoColor}patterns} of ports 2 and 3. 
The realized KPI of the set with modes $\lbrace 1,2,3\rbrace$ is about 4\,dB lower than that of the set containing modes $\lbrace 3,4,5\rbrace$. A dominant contribution of modes 4 and 5 to the far fields of ports 2 and 3 is therefore beneficial. 

If, additionally to modes 4 and 5, modes 1 and 2 had been selected, the number of desired modes belonging to $\Gamma^{(5)}$ would exceed the number of ports available for that representation. In this case, methods that accept higher port couplings or smaller bandwidths like \cite{Peitzmeier2020} can serve as a starting point for the design of independent ports for all four modes. Alternatively, the order of symmetry of the structure needs to be increased. Otherwise, sets that contain too many modes from an irreducible representation have to be excluded in the above procedure. In this example however, this issue does not occur. 


\begin{figure}
    \centering
    \includegraphics{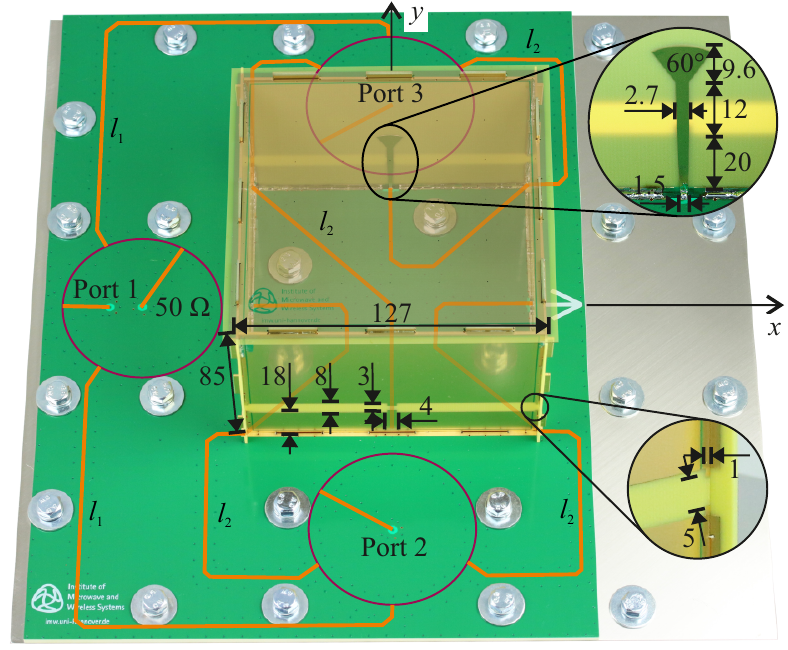}
    \caption{Depiction of the cuboid example antenna system, based on a photograph of the manufactured prototype. The image was edited to make the top part of the antenna appear semi-transparent. The strip lines on the inner layer of the multi layer feed network PCB (green) are highlighted in orange (impedance 50\,$\Omega$, width 1.635\,mm) and purple (impedance 70\,$\Omega$, width 0.793\,mm), respectively. The inner diameter of the ring hybrid couplers (purple) is 70.1\,mm. The lengths of the distribution lines (orange) are $l_1=262.4$\,mm and $l_2=252.4$\,mm, respectively. All dimensions are in mm. }
    \label{fig:prototypeDimensions}
\end{figure}
Since all ports have to utilize the same feed points, a feed network is required to distribute the signals between the ports and the feed points. The selected implementation consisting of three ring hybrid couplers is shown in Fig.~\ref{fig:prototypeDimensions}. The remaining port of the left hybrid coupler is terminated by a matched load. 
The feed network is implemented on a four-layer printed circuit board (PCB), identifiable by the green solder mask print in Fig.~\ref{fig:prototypeDimensions}. As substrate material, two layers of Rogers 4350B with a height of $1524\,\upmu$m each are utilized. 

The antenna itself is a hollow cuboid with a square footprint and the dimensions depicted in Fig.~\ref{fig:prototypeDimensions}. Its faces are made of 1\,mm thick, two-layer FR-4 PCBs. 
It should be highlighted that the antenna is not working as a cavity resonator, since the operating frequency range is significantly lower than the resonant frequency $f_\mathrm{res,TE110} \approx 1669\,\mathrm{MHz}$ of the first cavity mode. 
The $z$-position of the ideal delta gap feeds in Fig.~\ref{fig:cuboidCADPorts}, or equivalently, the $z$-position of the slot going around the antenna in Fig.~\ref{fig:prototypeDimensions}, determines to which modes from the respective irreducible representation the port patterns have the highest correlation. 
A compromise is chosen that improves input matching of port 1 while accepting a significant contribution of modes 1 and 2 to ports 2 and 3.

\subsection{Simulation Results}
\begin{figure}
\centering
\subfloat[Port 1]{
\includegraphics[]{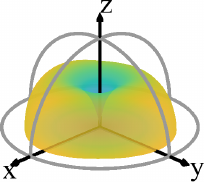}
\label{fig:antennaInfGNDP1}}
\hfil
\subfloat[Port 2]{
\includegraphics[]{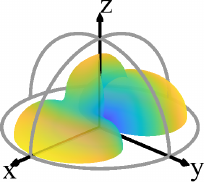}
\label{fig:antennaInfGNDP2}}
\hfil
\subfloat[Port 3]{
\includegraphics[]{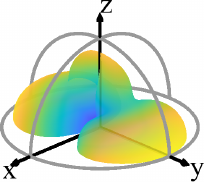}
\label{fig:antennaInfGNDP3}}
\hfil
\subfloat[]{
\includegraphics[]{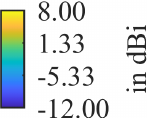}
\label{fig:antennaInfGNDcbar}}
\caption{Simulated $\Theta$-component of the directivity at 1060\,MHz for the three ports of the proposed antenna on an infinite ground plane. } \label{fig:antennaInfGNDFFs}
\end{figure}
The electromagnetic behavior of the presented demonstration antenna system on an infinite ground plane is determined by simulation in EmpireXPU. The simulation model includes the cuboid antenna as well as the complete feed network up to the ports' coaxial connectors.
The $\Theta$-components of the resulting far field directivities are shown in Fig.~\ref{fig:antennaInfGNDFFs}. As desired, the directivity of port 1 is very similar to that of mode 3, with the maximum directivity being 4.2\,dBi for the first and 4.0\,dBi for the latter, compare Fig.~\ref{fig:cuboidModes}. For ports 2 and 3 however, it is observed that the far fields are a superposition of the modes 1, 2, 4 and 5, as predicted above. The maximum directivity of the modes 4 and 5 is 6.5\,dBi, while it is approximately 1\,dB higher for the ports 2 and 3. In particular, a side lobe directed in $z$-direction is observed and a minimum near $\Theta=45^\circ$ is introduced. 

\begin{figure}
\centering
\subfloat[(a)]{
    \begin{tikzpicture}[baseline]
    \node[right, inner sep=0] (uMat) at (0,0) {
    \includegraphics[valign=c]{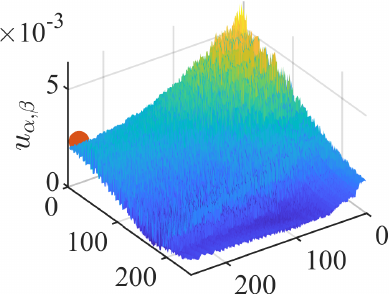} };
    \node (tstDoA) at (0.5, -1.4) {\footnotesize Test DoA};
    \node (refDoA) at (3.5, -1.4) {\footnotesize Ref. DoA};
    \end{tikzpicture}
\label{fig:antennaInfGNDC}}
\hfil
\subfloat[(b)]{
\includegraphics[valign=c]{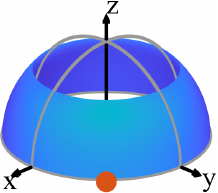}
\includegraphics[valign=c]{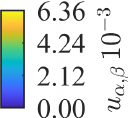}
\label{fig:antennaInfGNDCcol235}}
\caption{Simulated results for the ports of the example antenna on an infinite ground plane. The uncertainty matrix $\mathbf{U}$ for the three implemented ports is depicted in \protect\subref{fig:antennaInfGNDC}. 
The respective KPI is 32.7\,dB. The uncertainty vector $\mathbf{u}_\alpha$ for the reference DoA ($\Theta=90^\circ,\Phi=45^\circ$) is shown in \protect\subref{fig:antennaInfGNDCcol235}. } \label{fig:antennaInfGNDopt}
\end{figure}
The modifications of the far field patterns influence the uncertainty parameter of the demonstration antenna system, which is displayed in Fig.~\ref{fig:antennaInfGNDopt}. The general behavior of the modes is preserved in the sense that no ambiguity is observable in the uncertainty matrix in Fig.~\ref{fig:antennaInfGNDC}. However, the variation of the directivity w.\,r.\,t. the elevation in the investigated angular region is higher than for the modal far fields. Therefore, the uncertainty is higher for reference DoAs with low $\Theta$ angles, as indicated by the peak in Fig.~\ref{fig:antennaInfGNDopt}. 
Accordingly, the KPI of the simulated port setup is $\mathrm{KPI} \approx 32.7$\,dB, which is lower than the $\mathrm{KPI} \approx 34.2$\,dB of the pure Characteristic Modes $\lbrace 3,4,5\rbrace$ on the cuboid. Here, both KPIs are of the directivity type. 

\subsection{Measurement Results}
\begin{figure}
\centering
\includegraphics[width=0.5\columnwidth]{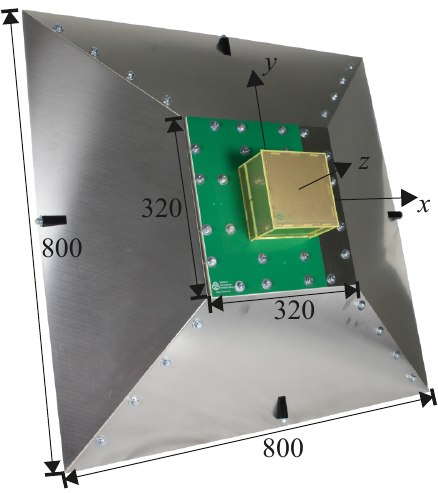}
\caption{Photograph and dimensions of the realized prototype of the cuboid antenna on an aircraft hull mock up. This is a truncated pyramid of 100\,mm height. Dimensions are in mm.} \label{fig:antennaPhoto}
\end{figure}

In order to allow measurements of the manufactured demonstrator in an anechoic chamber, the infinite ground plane, which was introduced as a model for the hull of an aircraft, needs to be replaced by a mock up of finite dimensions. For this, a truncated pyramid is chosen. The final setup is shown in Fig.~\ref{fig:antennaPhoto}. 
\begin{figure}
\centering
\includegraphics[valign=t]{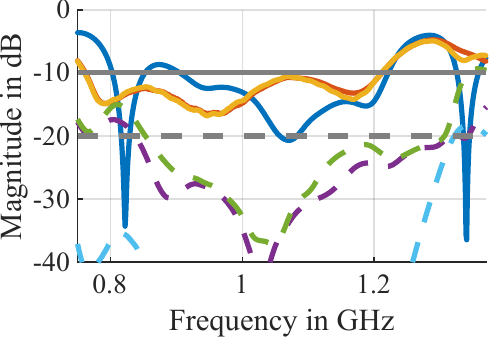} 
\hfil
\includegraphics[valign=t]{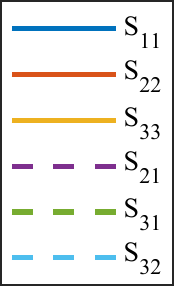} 
\caption{Measured scattering parameters of the fabricated demonstrator. The input reflections (solid lines) are below $-10$\,dB and the transmissions (dashed lines) below $-20$\,dB over a large bandwidth of at least 28\%. } \label{fig:antennaSparams}
\end{figure}
The measured scattering parameters are found in Fig.~\ref{fig:antennaSparams}. It is observed that the input reflections at the investigated ACAS frequencies 1030\,MHz and 1090\,MHz are below $-10$\,dB for all ports. The transmissions between the ports are far below $-20$\,dB, showing that the ports are indeed decorrelated. 
\begin{figure}
\centering
\subfloat[Port 1]{
\includegraphics[valign=c]{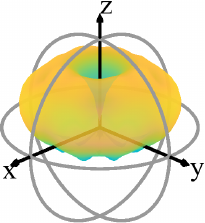}
\label{fig:antennaLargeGNDP1}}
\hfil
\subfloat[Port 2]{
\includegraphics[valign=c]{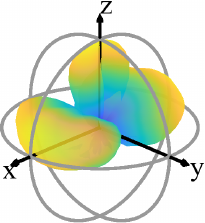}
\label{fig:antennaLargeGNDP2}}
\hfil
\subfloat[Port 3]{
\includegraphics[valign=c]{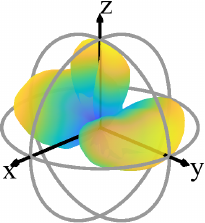}
\label{fig:antennaLargeGNDP3}}
\hfil
\subfloat[]{
\includegraphics[valign=c]{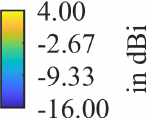}
\label{fig:antennaLargeGNDcbar}}
\caption{Measured $\Theta$-component of the realized gain at 1060\,MHz of the three ports of the proposed antenna on a ground plane mock up. } \label{fig:antennaLargeGNDFFs}
\end{figure}
The measured realized gain patterns of the demonstrator are depicted in Fig.~\ref{fig:antennaLargeGNDFFs}. It is found that the maximum of the far field of port 1 moves away from the ground plane mock up, while the peaks of ports 2 and 3 maintain their position, compared to the simulation results. This effect is attributed to the size and shape of the ground plane mock up and matches simulation results for this case. It highlights that the integration scenario has a great impact on the behavior of the antenna system. 
\begin{figure}
\centering
\subfloat[(a)]{
    \begin{tikzpicture}[baseline]
    \node[right, inner sep=0] (uMat) at (0,0) {
    \includegraphics[valign=c]{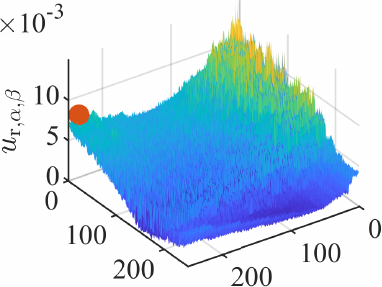} };
    \node (tstDoA) at (0.5, -1.4) {\footnotesize Test DoA};
    \node (refDoA) at (3.5, -1.4) {\footnotesize Ref. DoA};
    \end{tikzpicture}
\label{fig:antennaLargeGNDC}}
\hfil
\subfloat[(b)]{
\includegraphics[valign=c]{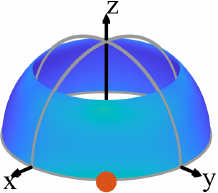}
\includegraphics[valign=c]{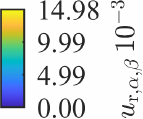}
\label{fig:antennaLargeGNDCcol235}}
\caption{Measured results for the ports of the example antenna on a ground plane mock up. The realized uncertainty matrix $\mathbf{U}_\mathrm{realized}$ for the three implemented ports is depicted in \protect\subref{fig:antennaLargeGNDC}. 
The respective KPI is 29.0\,dB. The realized uncertainty vector $\mathbf{u}_\mathrm{realized,\alpha}$ for the reference DoA ($\Theta=90^\circ,\Phi=45^\circ$) is shown in \protect\subref{fig:antennaLargeGNDCcol235}. } \label{fig:antennaLargeGNDopt}
\end{figure}
Again, the method proposed in this work provides a useful tool for the analysis of the consequences of this behavior {\color{typoColor}w.\,r.\,t.} direction finding. The uncertainty parameters for the measured far fields are shown in Fig.~\ref{fig:antennaLargeGNDopt}. As expected, no ambiguities are observed, while an increase of the uncertainty for both low and high elevation angles is seen in the uncertainty matrix and the example uncertainty vector. 
This effect is mainly attributed to the limited size of the ground {\color{typoColor}plane mock up}. 

\section{Conclusion} \label{sec:conclusion}
A multi-step evaluation procedure for arbitrary multi-port direction finding antennas is presented. 
In contrast to established methods like the CRB or MUSIC, it works without the need to consider noise, making it simpler and more intuitive. It thereby allows the user to gain more insight into the physical behavior of the antenna system. Also, there are fewer parameters to vary, improving comparability between different designs. 

The proposed procedure is utilized as a design guidance for a multi-mode multi-port antenna system. Starting with a set of three Characteristic Modes of a hemisphere on an infinite ground plane, the insight gained by the proposed analysis procedure is transferred to three CMs on a smaller cuboid. A set of uncorrelated ports is derived from the selected CMs and simulated. A demonstrator is measured and the influence of a finite ground plate, mismatch and losses are accounted for. Using the proposed visualizations, it is verified that during this transfer from ideal modal parameters to real world port measurements, undesired effects like ambiguities are ruled out. 


\appendix  
A derivation of the estimated incident far fields and the incident field modal expansion coefficients is given in the following with the help of \cite{Grundmann2021a, Capek2023}. 
We assume $\mathrm{e}^{\mathrm{j}\omega t}$ time dependency with $\omega$ being the angular frequency and $t$ the time. The electric field of a $\gamma'$ polarized plane wave traveling in $-\mathbf{e}_\mathrm{r}'$ direction, which means it is incident from the $\mathbf{e}_\mathrm{r}'$ direction, see Fig.~\ref{fig:introduction}a, is defined as \cite{Capek2023}: 
\begin{equation}
    \mathbf{E}_\mathrm{inc,\gamma'}(\mathbf{e}_\mathrm{r}',\mathbf{r}) 
    = E_{0,\gamma'}\mathbf{e}_{\gamma'} \mathrm{e}^{-\mathrm{j}k_0 (-\mathbf{e}_\mathrm{r}') \cdot\mathbf{r}}
    = E_{0,\gamma'}\mathbf{e}_{\gamma'} \mathrm{e}^{\mathrm{j}k_0 \mathbf{e}_\mathrm{r}'\cdot\mathbf{r}}  ~.
\end{equation}
Here, $\mathbf{e}_{\gamma'}$ is the unit vector in the direction of the $\gamma'$ polarization, $E_{0,\gamma'}$ the complex amplitude of that component and $\mathbf{r}$ the observation position. 

The plane wave is incident on a PEC surface, discretized using Rao-Wilton-Glisson (RWG) basis functions \cite{Rao1982}. 
The electric field weight at the $i$-th RWG $\boldsymbol{\psi}_i(\mathbf{r})$ becomes: 
\begin{align}
    V_{\gamma',i}(\mathbf{e}_\mathrm{r}') 
    &= E_{0,\gamma'}\mathbf{e}_{\gamma'} \int\limits_{{\mathbb{R}}^3} \boldsymbol{\psi}_i(\mathbf{r}) \cdot \mathrm{e}^{\mathrm{j}k_0 \mathbf{e}_\mathrm{r}'\cdot\mathbf{r}} \mathrm{d}V ~. \label{eq:V2}
\end{align}
The $\gamma$-component of the radiated far field in direction $\mathbf{e}_\mathrm{r}$, originating from a current $\mathbf{I}$, is defined as: 
\begin{equation} \label{eq:F}
    F_{\gamma}(\mathbf{e}_\mathrm{r}) = \mathbf{K}_\gamma(\mathbf{e}_\mathrm{r}) \mathbf{I}~.
\end{equation}
Here, $\mathbf{K}_\gamma(\mathbf{e}_\mathrm{r})$ is the far field row vector that maps the current at the $i$-th RWG to the $\gamma$-polarized far field in $\mathbf{e}_\mathrm{r}$ direction. The entries of this row vector are defined as \cite{Capek2023}
\begin{equation}\label{eq:K}
    K_{\gamma,i} (\mathbf{e}_\mathrm{r}) = -\mathrm{j}\frac{Z_0 k_0}{4\pi} \mathbf{e}_{\gamma} \int\limits_{{\mathbb{R}}^3} \boldsymbol{\psi}_i(\mathbf{r}) \cdot \mathrm{e}^{\mathrm{j}k_0 \mathbf{e}_\mathrm{r}\cdot\mathbf{r}} \mathrm{d}V ~,
\end{equation}
where $Z_0$ is the free space wave impedance. 
By evaluating the far field row vector (\ref{eq:K}) in the direction $\mathbf{e}_\mathrm{r}'$ the plane wave is incident from and comparing it with (\ref{eq:V2}), we obtain \cite{Capek2023}
\begin{equation} \label{eq:VfromK}
    \mathbf{V}_{\gamma'}(\mathbf{e}_\mathrm{r}') = \mathrm{j}\frac{4\pi}{Z_0 k_0} E_{0,\gamma'} \mathbf{K}_{\gamma'}^\mathrm{T} (\mathbf{e}_\mathrm{r}') ~,
\end{equation}
with the entries of the column vector $\mathbf{V}_{\gamma'}(\mathbf{e}_\mathrm{r}')$ being the results of (\ref{eq:V2}), while $(\cdot)^\mathrm{T}$ is the transpose. 

Next, the $\gamma$-component of the estimated incident far field $F_\mathrm{inc,est,\gamma',\gamma} (\mathbf{e}_\mathrm{r}', \mathbf{e}_\mathrm{r})$ is calculated from the Characteristic Mode far fields following (\ref{eq:FincestSum}). 
Here, the incident field modal expansion coefficient $c_{\gamma',n} (\mathbf{e}_\mathrm{r}')$ is determined by \cite{Grundmann2021a}
\begin{equation} \label{eq:cnFromT}
    c_{\gamma',n} (\mathbf{e}_\mathrm{r}') = \frac{1}{2 T_{n,n}} a_{\gamma',n} (\mathbf{e}_\mathrm{r}') ~,
\end{equation}
where the modal weighting coefficient for scattered fields $a_{\gamma',n} (\mathbf{e}_\mathrm{r}')$ is given in (\ref{eq:sctFnSuperposition})
and $T_{n,n}$ are the diagonal entries of the transition matrix $\mathbf{T}$, which relates an incident field to a scattered field \cite{Gustafsson2022}:  
\begin{equation} \label{eq:tn}
    T_{n,n} = t_n = -\frac{1}{1+\mathrm{j}\lambda_n} ~.
\end{equation}
With (\ref{eq:sctFnSuperposition}) and (\ref{eq:tn}), (\ref{eq:cnFromT}) becomes \cite{Grundmann2021a}: 
\begin{equation} \label{eq:cn}
    c_{\gamma',n} (\mathbf{e}_\mathrm{r}') = -\frac{1}{4} \mathbf{I}_n^\mathrm{H} \mathbf{V}_\mathrm{\gamma'} (\mathbf{e}_\mathrm{r}') ~.
\end{equation}
If we insert (\ref{eq:VfromK}) in (\ref{eq:cn}) and utilize that $\mathbf{I}_n$ is real, we obtain: 
\begin{equation}
    c_{\gamma',n} (\mathbf{e}_\mathrm{r}') 
    = - \mathrm{j}\frac{\pi}{Z_0 k_0} E_{0,\gamma'} (\mathbf{K}_{\gamma'} (\mathbf{e}_\mathrm{r}') \mathbf{I}_n)^\mathrm{T} ~.
\end{equation}
With (\ref{eq:F}), we finally arrive at: 
\begin{equation} \label{eq:cn_final}
    c_{\gamma',n} (\mathbf{e}_\mathrm{r}') = - \mathrm{j}\frac{\pi}{Z_0 k_0} E_{0,\gamma'} F_{\gamma',n}(\mathbf{e}_\mathrm{r}') ~.
\end{equation}
Comparing (\ref{eq:cn_final}) to Fig.~\ref{fig:FarFieldMeasMat} and replacing the port index $p$ with the mode index $n$, {\color{typoColor}it is found} that the entries in the measurement matrix are proportional to the incident field modal expansion coefficients. The proportionality factor is identical for all entries. Therefore, the correlation (\ref{eq:DoACorrelations}) is {\color{typoColor}equivalent} to (\ref{eq:rhoFromC}). 

\ifCLASSOPTIONcaptionsoff
  \newpage
\fi



\bibliographystyle{IEEEtran}
\bibliography{IEEEabrv,literatur_master360}

\begin{thebibliography}{10}
\providecommand{\url}[1]{#1}
\csname url@samestyle\endcsname
\providecommand{\newblock}{\relax}
\providecommand{\bibinfo}[2]{#2}
\providecommand{\BIBentrySTDinterwordspacing}{\spaceskip=0pt\relax}
\providecommand{\BIBentryALTinterwordstretchfactor}{4}
\providecommand{\BIBentryALTinterwordspacing}{\spaceskip=\fontdimen2\font plus
\BIBentryALTinterwordstretchfactor\fontdimen3\font minus
  \fontdimen4\font\relax}
\providecommand{\BIBforeignlanguage}[2]{{%
\expandafter\ifx\csname l@#1\endcsname\relax
\typeout{** WARNING: IEEEtran.bst: No hyphenation pattern has been}%
\typeout{** loaded for the language `#1'. Using the pattern for}%
\typeout{** the default language instead.}%
\else
\language=\csname l@#1\endcsname
\fi
#2}}
\providecommand{\BIBdecl}{\relax}
\BIBdecl

\bibitem{Wild2021}
T.~Wild, V.~Braun, and H.~Viswanathan, ``Joint design of communication and
  sensing for beyond 5g and 6g systems,'' \emph{IEEE Access}, vol.~9, pp.
  30\,845--30\,857, 2021.

\bibitem{ICAO2006}
\emph{Airborne Collision Avoidance System (ACAS) Manual}, 1st~ed.,
  International Civil Aviation Organization (ICAO), 2006.

\bibitem{Balanis2016}
C.~A. Balanis, \emph{Antenna Theory: Analysis and Design}, 4th~ed.\hskip 1em
  plus 0.5em minus 0.4em\relax Wiley, Feb. 2016.

\bibitem{Harrington1971a}
R.~{Harrington} and J.~{Mautz}, ``Theory of characteristic modes for conducting
  bodies,'' \emph{IEEE Transactions on Antennas and Propagation}, vol.~19,
  no.~5, pp. 622--628, Sep. 1971.

\bibitem{Peitzmeier2019}
N.~{Peitzmeier} and D.~{Manteuffel}, ``Upper bounds and design guidelines for
  realizing uncorrelated ports on multimode antennas based on symmetry analysis
  of characteristic modes,'' \emph{IEEE Transactions on Antennas and
  Propagation}, vol.~67, no.~6, pp. 3902--3914, June 2019.

\bibitem{Peitzmeier2022}
N.~Peitzmeier, T.~Hahn, and D.~Manteuffel, ``Systematic {Design} of {Multimode}
  {Antennas} for {MIMO} {Applications} by {Leveraging} {Symmetry},'' \emph{IEEE
  Transactions on Antennas and Propagation}, vol.~70, no.~1, pp. 145--155, Jan.
  2022.

\bibitem{Manteuffel2022}
D.~Manteuffel, F.~H. Lin, T.~Li, N.~Peitzmeier, and Z.~N. Chen,
  ``Characteristic mode-inspired advanced multiple antennas: Intuitive insight
  into element-, interelement-, and array levels of compact large arrays and
  metantennas,'' \emph{IEEE Antennas and Propagation Magazine}, vol.~64, no.~2,
  pp. 49--57, April 2022.

\bibitem{Ma2019}
R.~{Ma} and N.~{Behdad}, ``Design of platform-based hf direction-finding
  antennas using the characteristic mode theory,'' \emph{IEEE Transactions on
  Antennas and Propagation}, vol.~67, no.~3, pp. 1417--1427, March 2019.

\bibitem{Poehlmann2019}
R.~{P\"ohlmann}, S.~A. {Almasri}, S.~{Zhang}, T.~{Jost}, A.~{Dammann}, and
  P.~A. {Hoeher}, ``On the potential of multi-mode antennas for
  direction-of-arrival estimation,'' \emph{IEEE Transactions on Antennas and
  Propagation}, vol.~67, no.~5, pp. 3374--3386, May 2019.

\bibitem{AlkubtiAlmasri2019}
S.~{Alkubti Almasri}, R.~{P\"ohlmann}, N.~{Doose}, P.~A. {Hoeher}, and
  A.~{Dammann}, ``Modeling aspects of planar multi-mode antennas for
  direction-of-arrival estimation,'' \emph{IEEE Sensors Journal}, vol.~19,
  no.~12, pp. 4585--4597, June 2019.

\bibitem{Ren2021}
K.~Ren, R.~Ma, and N.~Behdad, ``Performance-enhancement of platform-based, hf
  direction-finding systems using dynamic mode selection,'' \emph{IEEE Open
  Journal of Antennas and Propagation}, 2021.

\bibitem{Grundmann2021}
L.~Grundmann, N.~Peitzmeier, and D.~Manteuffel, ``Investigation of direction of
  arrival estimation using characteristic modes,'' in \emph{2021 15th European
  Conference on Antennas and Propagation (EuCAP)}, March 2021.

\bibitem{Grundmann2022}
L.~Grundmann and D.~Manteuffel, ``Selecting characteristic modes in multi-mode
  direction finding antenna design by using reconstructed incident fields,'' in
  \emph{2022 16th European Conference on Antennas and Propagation (EuCAP)},
  March 2022.

\bibitem{Weiss1991}
A.~Weiss and B.~Friedlander, ``Performance analysis of diversely polarized
  antenna arrays,'' \emph{IEEE Transactions on Signal Processing}, vol.~39,
  no.~7, pp. 1589--1603, July 1991.

\bibitem{Hua1991}
Y.~Hua, T.~Sarkar, and D.~Weiner, ``An l-shaped array for estimating 2-d
  directions of wave arrival,'' \emph{IEEE Transactions on Antennas and
  Propagation}, vol.~39, no.~2, pp. 143--146, Feb 1991.

\bibitem{Nordebo2006}
S.~{Nordebo}, M.~{Gustafsson}, and J.~{Lundback}, ``Fundamental limitations for
  doa and polarization estimation with applications in array signal
  processing,'' \emph{IEEE Transactions on Signal Processing}, vol.~54, no.~10,
  pp. 4055--4061, Oct 2006.

\bibitem{Jackson2015}
B.~R. Jackson, S.~Rajan, B.~J. Liao, and S.~Wang, ``Direction of arrival
  estimation using directive antennas in uniform circular arrays,'' \emph{IEEE
  Transactions on Antennas and Propagation}, vol.~63, no.~2, pp. 736--747, Feb
  2015.

\bibitem{Pralon2017}
M.~G. Pralon, G.~Del~Galdo, M.~Landmann, M.~A. Hein, and R.~S. Thomä,
  ``Suitability of compact antenna arrays for direction-of-arrival
  estimation,'' \emph{IEEE Transactions on Antennas and Propagation}, vol.~65,
  no.~12, pp. 7244--7256, Dec 2017.

\bibitem{Li2020}
S.~Li, Y.~Liu, L.~You, and W.~Wang, ``Optimal three-dimensional antenna array
  for direction finding with geometric constraint,'' \emph{IEEE Access},
  vol.~8, pp. 31\,948--31\,956, 2020.

\bibitem{Ma2022}
R.~Ma and N.~Behdad, ``A spatially confined, platform-based hf direction
  finding array,'' \emph{IEEE Transactions on Antennas and Propagation},
  vol.~70, no.~2, pp. 1298--1308, Feb 2022.

\bibitem{Schmidt1986}
R.~{Schmidt}, ``Multiple emitter location and signal parameter estimation,''
  \emph{IEEE Transactions on Antennas and Propagation}, vol.~34, no.~3, pp.
  276--280, March 1986.

\bibitem{DeB.Gripp2017}
T.~A. {De B. Gripp}, B.~M. {Fabiani}, E.~S. {Silveira}, and D.~C. {Nascimento},
  ``Design of a microstrip antenna array with polarization diversity for doa
  application,'' in \emph{2017 SBMO/IEEE MTT-S International Microwave and
  Optoelectronics Conference (IMOC)}, Aug 2017.

\bibitem{Kabiri2020}
S.~Kabiri, E.~Kornaros, and F.~De~Flaviis, ``Tightly coupled array design based
  on phase center contour for indoor direction findings in harsh
  environments,'' \emph{IEEE Transactions on Antennas and Propagation},
  vol.~68, no.~4, pp. 2698--2713, April 2020.

\bibitem{Park2021}
C.~J. Park, C.~N. Pearce, A.~Ackie, F.~Vassallo, and K.~J. Duncan,
  ``Dynamically reconfigurable direction-finding antenna array for unmanned
  arial systems,'' in \emph{2021 IEEE 21st Annual Wireless and Microwave
  Technology Conference (WAMICON)}, April 2021.

\bibitem{Kataria2019}
C.~Y. Kataria, G.~X. Gao, and J.~T. Bernhard, ``Design of a compact hemispiral
  gps antenna with direction finding capabilities,'' \emph{IEEE Transactions on
  Antennas and Propagation}, vol.~67, no.~5, pp. 2878--2885, May 2019.

\bibitem{Grundmann2021a}
L.~Grundmann and D.~Manteuffel, ``Using characteristic modes for determining
  the incident field in a scattering problem,'' in \emph{2021 IEEE
  International Symposium on Antennas and Propagation and USNC-URSI Radio
  Science Meeting (APS/URSI)}, Dec 2021, pp. 855--856.

\bibitem{Bailey2012}
M.~C. Bailey, T.~G. Campbell, C.~J. Reddy, R.~L. Kellogg, and P.~Nguyen,
  ``Compact wideband direction-finding antenna,'' \emph{IEEE Antennas and
  Propagation Magazine}, vol.~54, no.~6, pp. 44--68, December 2012.

\bibitem{Capek2023}
M.~Capek, J.~Lundgren, M.~Gustafsson, K.~Schab, and L.~Jelinek,
  ``Characteristic mode decomposition using the scattering dyadic in arbitrary
  full-wave solvers,'' \emph{IEEE Transactions on Antennas and Propagation},
  vol.~71, no.~1, pp. 830--839, Jan 2023.

\bibitem{Pozar2003}
D.~Pozar, ``A relation between the active input impedance and the active
  element pattern of a phased array,'' \emph{IEEE Transactions on Antennas and
  Propagation}, vol.~51, no.~9, pp. 2486--2489, Sep. 2003.

\bibitem{Chen2018}
Y.~Chen, K.~Schab, M.~Čapek, M.~Mašek, B.~K. Lau, H.~Aliakbari, Y.~Haykir,
  Q.~Wu, W.~Strydom, N.~Peitzmeier, M.~Jovicic, S.~Genovesi, and F.~A.
  Dicandia, ``Benchmark problem definition and cross-validation for
  characteristic mode solvers,'' in \emph{12th European Conference on Antennas
  and Propagation (EuCAP 2018)}, April 2018.

\bibitem{Safin2013}
E.~Safin and D.~Manteuffel, ``Reconstruction of the characteristic modes on an
  antenna based on the radiated far field,'' \emph{IEEE Transactions on
  Antennas and Propagation}, vol.~61, no.~6, pp. 2964--2971, June 2013.

\bibitem{Gagarinov2017}
P.~Gagarinov, ``Spheretri,'' \url{https://github.com/pgagarinov/spheretri},
  2017.

\bibitem{empirexpu}
\BIBentryALTinterwordspacing
\emph{EMPIRE XPU Software}. [Online]. Available: \url{https://www.empire.de}
\BIBentrySTDinterwordspacing

\bibitem{Gazzah2006}
H.~Gazzah and S.~Marcos, ``Cramer-rao bounds for antenna array design,''
  \emph{IEEE Transactions on Signal Processing}, vol.~54, no.~1, pp. 336--345,
  Jan 2006.

\bibitem{Cornwell1997}
J.~F. Cornwell, \emph{\BIBforeignlanguage{en}{Group {Theory} in {Physics}: {An}
  {Introduction}}}.\hskip 1em plus 0.5em minus 0.4em\relax Academic Press, Jul.
  1997.

\bibitem{Peitzmeier2020}
N.~Peitzmeier and D.~Manteuffel, ``Systematic design method for asymmetric
  multiport antennas based on characteristic modes,'' in \emph{2020 14th
  European Conference on Antennas and Propagation (EuCAP)}, March 2020.

\bibitem{Rao1982}
S.~{Rao}, D.~{Wilton}, and A.~{Glisson}, ``Electromagnetic scattering by
  surfaces of arbitrary shape,'' \emph{IEEE Transactions on Antennas and
  Propagation}, vol.~30, no.~3, pp. 409--418, 1982.

\bibitem{Gustafsson2022}
M.~Gustafsson, L.~Jelinek, K.~Schab, and M.~Capek, ``Unified theory of
  characteristic modes—part i: Fundamentals,'' \emph{IEEE Transactions on
  Antennas and Propagation}, vol.~70, no.~12, pp. 11\,801--11\,813, Dec 2022.

\end{thebibliography}

\begin{IEEEbiography}[{\includegraphics[width=1in,height=1.25in,clip,keepaspectratio]{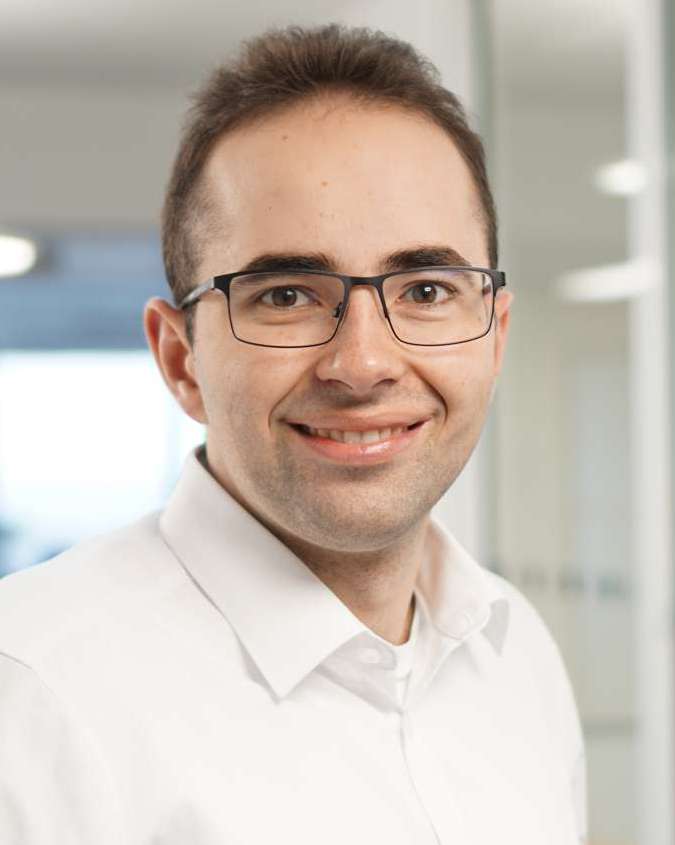}}]{Lukas Grundmann} (Graduate Student Member, IEEE) was born in 1994 in Hoya, Germany. He received the B.Sc. and M.Sc. degrees in electrical engineering from Leibniz University Hannover, Hannover, Germany, in 2017 and 2019, respectively. He is currently a Research Assistant with the Institute of Microwave and Wireless Systems, Leibniz University Hannover. His current research focuses on modal expansion techniques, such as spherical wave functions and characteristic modes, and their applications to antenna development. 
\end{IEEEbiography}

\begin{IEEEbiography}[{\includegraphics[width=1in,height=1.25in,clip,keepaspectratio]{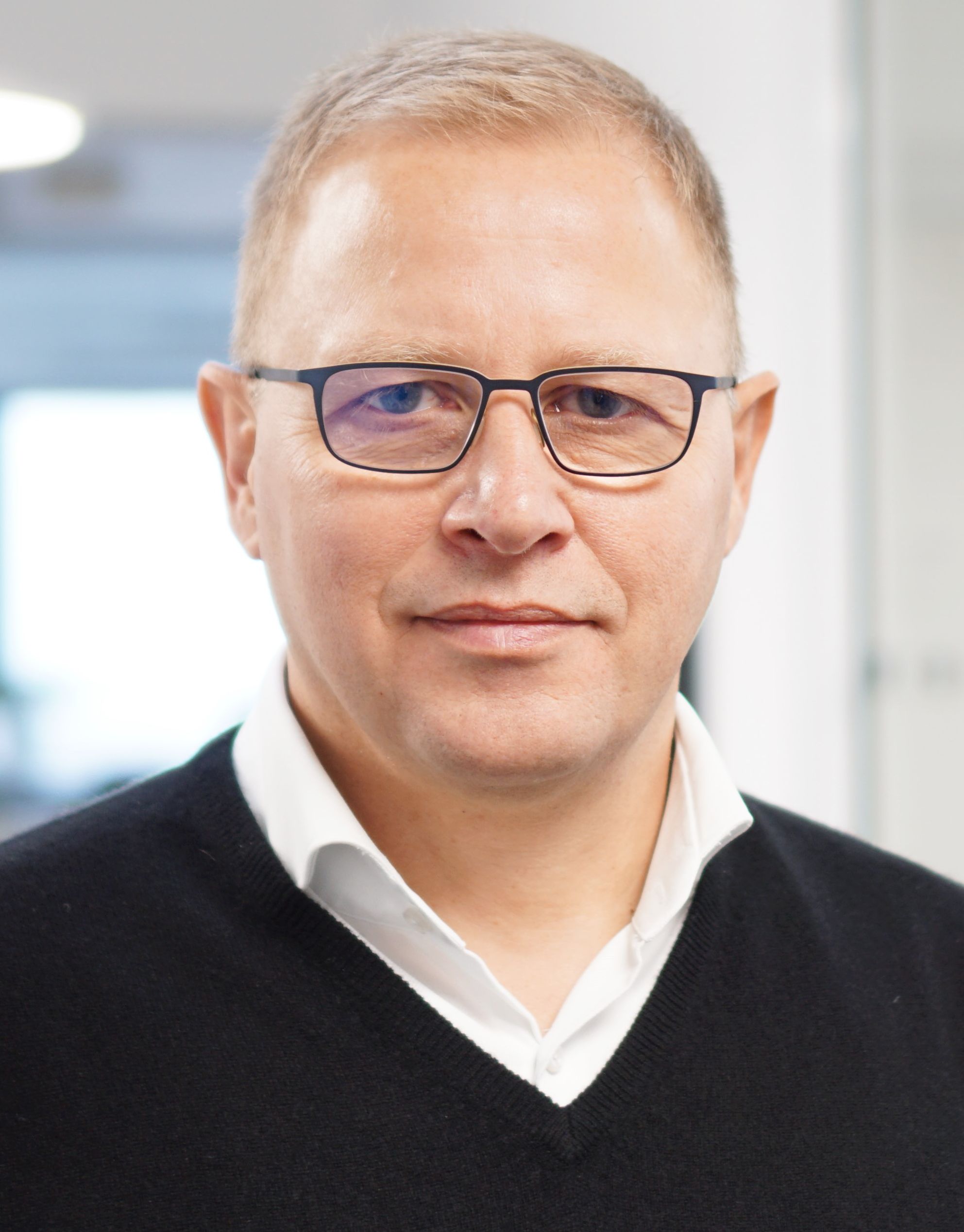}}]{Dirk Manteuffel} (Member, IEEE) was born in Issum, Germany, in 1970. He received the Dipl.-Ing. and Dr.-Ing. degrees in electrical engineering from
the University of Duisburg–Essen, Duisburg, Germany, in 1998 and 2002, respectively.
From 1998 to 2009, he was with IMST, Kamp-Lintfort, Germany. As a Project Manager, he was responsible for industrial antenna development and advanced projects in the field of antennas and electromagnetic (EM) modeling. From 2009 to 2016, he was a Full Professor of wireless communications at Christian-Albrechts-University, Kiel, Germany. Since June 2016, he has been a Full Professor and the Executive Director of the Institute of Microwave and Wireless Systems, Leibniz University Hannover, Hannover, Germany. His research interests include electromagnetics, antenna integration and EM modeling for mobile communications and biomedical applications.
Dr. Manteuffel was a director of the European Association on Antennas and Propagation from 2012 to 2015. He served on the Administrative Committee (AdCom) of IEEE Antennas and Propagation Society from 2013 to 2015 and as an Associate Editor of the IEEE Transactions on Antennas and Propagation from 2014 to 2022. Since 2009 he has been an appointed member of the committee "Antennas" of the German VDI-ITG. 

\end{IEEEbiography}

\end{document}